\documentclass{mn2e}
\usepackage{epsfig}
\usepackage{graphicx}

\title{Far-IR Detection Limits I: \\ Sky Confusion Due to Galactic Cirrus}
\author[W.-S.~Jeong et al.]
    {Woong-Seob~Jeong,$^1$\thanks{Woong-Seob~Jeong (jeongws@astro.snu.ac.kr)}
     Hyung Mok~Lee,$^1$ Soojong~Pak,$^2$ Takao~Nakagawa,$^3$
     \newauthor Suk Minn~Kwon,$^{3,4}$ Chris~P.~Pearson$^{3,5}$ and Glenn J.~White$^5$ \\
    $^1$ Astronomy Program in Graduate School of Earth and Environmental Sciences,
    Seoul National University, \\ ~~~Shillim-Dong Kwanak-Gu, Seoul 151-742, South Korea \\
    $^2$ Korea Astronomy Observatory, Whaam-Dong, Youseong-Gu, Taejeon 305-348, South Korea \\
    $^3$ Institute of Space and Astronautical Science, Japan Aerospace Exploration
    Agency, \\ ~~Yoshinodai 3-1-1, Sagamihara, Kanagawa 229-8510, Japan \\
    $^4$ Department of Science Education, Kangwon National
    University, Hyoja-Dong, Chunchon-Si, Kangwon-Do 200-701, South Korea \\
    $^5$ Center for Astrophysics and Planetary Science, University of Kent, Canterbury, Kent, CT2 7NR, England}

\date{Accepted . Received  ;  in original form 2003 January }

\pagerange{\pageref{firstpage}--\pageref{lastpage}} \pubyear{----}

\def\LaTeX{L\kern-.36em\raise.3ex\hbox{a}\kern-.15em
    T\kern-.1667em\lower.7ex\hbox{E}\kern-.125emX}

\begin{document}

\label{firstpage}

\maketitle

\begin{abstract}
Fluctuations in the brightness of the background radiation can lead to confusion
with real point sources. Such background emission confusion will be important for
infrared observations with relatively large beam sizes since the amount of
fluctuation tends to increase with angular scale. In order to quantitively assess
the effect of the background emission on the detection of point sources for current
and future far-infrared observations by space-borne missions such as
\textit{Spitzer}, \textit{ASTRO-F}, \textit{Herschel} and \textit{SPICA}, we have
extended the Galactic emission map to higher angular resolution than the currently
available data. Using this high resolution map, we estimate the sky confusion noise
due to the emission from interstellar dust clouds or cirrus, based on fluctuation
analysis and detailed photometry over realistically simulated images. We find that
the confusion noise derived by simple fluctuation analysis agrees well with the
result from realistic simulations. Although the sky confusion noise becomes dominant
in long wavelength bands ($> 100~\mu$m) with 60 -- 90cm aperture missions, it is
expected to be two order of magnitude smaller for the next generation space missions
with larger aperture sizes such as \textit{Herschel} and \textit{SPICA}.
\end{abstract}


\section{INTRODUCTION}

The detection of faint sources in far IR can be greatly affected by the amount and
structure of the background radiation. The main source of background radiation in
far IR is the smooth component of the Galactic emission, known as cirrus emission.
The amount of emission manifests itself as photon noise whose fluctuations follow
Poisson statistics. In addition, any brightness fluctuation at scales below the beam
size could cause confusion with real point sources. The cirrus emission was
discovered by the Infrared Astronomy Satellite (\textit{IRAS}) \cite{low84}, and is
thought to be due to radiatively heated interstellar dust in irregular clouds of
wide ranges of spatial scales. The cirrus emission peaks at far-IR wavelengths but
was detected in all four \textit{IRAS} bands at 12, 25, 60, and 100 $\mu$m (Helou \&
Beichman 1990, hereafter HB0). The brightness of cirrus emission depends upon the
Galactic latitude and is significant for wavelengths longer than 60 $\mu$m. The
cirrus emission, which is the main source of background radiation in far-IR, causes
an uncertainty in the determination of source fluxes use its brightness varies from
place to place. The accurate determination of observational detection limits
requires a knowledge of the cirrus emission as a function of position on the sky.
The other important factor affecting the source detection is the source confusion
which mainly depends upon the telescope beam size and the source distribution
itself. The effects resulting from a combination of the sky confusion and the source
confusion will be discussed in depth in the forthcoming paper [Jeong et al. 2004c
\shortcite{jeong04c}, in preparation], and we concentrate on the effect of sky
confusion in the present paper.

There have been realistic estimations of the sky confusion from observational data
from \textit{IRAS} and the Infrared Space Observatory (\textit{ISO}) (Gautier et al.
1992; HB90; Herbstmeier et al. 1998; Kiss et al. 2001). However, the resolution of
the data from \textit{IRAS} and \textit{ISO} is not sufficient to the application to
larger missions planned in future. Many valuable data in the far-IR wavelength range
will be available within or around this decade by a multitude of IR space projects
such as \textit{Spitzer} \cite{gall03}, \textit{ASTRO-F}
\cite{mura98,shib00,naka01,pearson04}, Herschel Space Observatory (\textit{HSO})
\cite{pilb03,poglit03} and the Space Infrared Telescope for Cosmology and
Astrophysics (\textit{SPICA}) \cite{naka04}. Since these instruments will observe
the sky with high sensitivities and high angular resolution, it is necessary to
understand the factors determining their detection limits.

The purpose of the present paper is to investigate the effects of cirrus emission on
the detection of faint point sources in highly sensitive future infrared
observations. Based on the measured power spectrum and the spectral energy
distribution models of the dust emission over the entire sky, we generate the dust
map with higher spatial resolution in various relevant wavelength bands by
extrapolating the power spectrum to small scales.

This paper is organized as follows. In Section \ref{sec:sky_conf}, we briefly
describe the sky confusion noise due to sky brightness fluctuations. In Section
\ref{sec:gen_dmap}, the high angular resolution realization of Galactic dust
emission in various IR bands is presented. Based upon the specifications of each IR
mission, we estimate the sky confusion noise by using simple fluctuation analysis in
Section \ref{sec:stat_analy_scn}. We compare estimated detection limits based on
fluctuation analysis with the results based on the photometry on realistically
simulated data in Section \ref{sec:scn_phot}. Our conclusions are summarised in
Section \ref{sec:summary}.

\section{CONFUSION DUE TO SKY FLUCTUATION}\label{sec:sky_conf}

\begin{figure}
    \centering \centerline{
    \psfig{figure=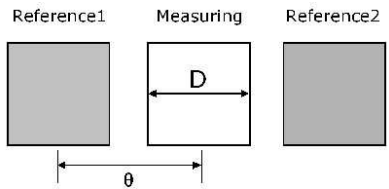,height=2.5cm} }
    \caption{Schematic outline of the reference aperture configurations
    for two symmetrically placed circular apertures (Gautier et al. 1992).}
    \label{fig_ref_aper}
\end{figure}

Measuring the brightness of sources involves subtracting the sky background derived
from the well-defined reference. The fluctuations in the surface brightness of
extended structure on similar scales to the resolution of the telescope and
instrument beam can produce spurious events that can be easily mistaken for genuine
point sources. This is because the source detection is usually simply accomplished
from the difference in signal between the on-source position and some background
position. Therefore sky confusion noise due to the sky brightness fluctuations,
$N(\theta)$, is defined as (HB90; Gautier et al. 1992):
\begin{equation}
    N(\theta) = \Omega \sqrt{S(\theta)}, \label{eqn_noise}
\end{equation}
where $\Omega$ is the solid angle of the measuring aperture, $\theta$ is the angular
separation between the target and reference sky positions, and $S(\theta)$ is the
second order structure function, which is defined as \cite{gautier92}:
\begin{equation}
S(\theta) = \left\langle \left | I(x) - \frac{I(x - \theta) + I(x + \theta)}{2}
\right |^2 \right\rangle _x , \label{eqn_strno}
\end{equation}
where $I$ is the sky brightness, $x$ is the location of the target, and $\langle~
\rangle$ represents the average taken over the whole map. For the configuration of
two symmetrically placed reference apertures, see Fig. \ref{fig_ref_aper}.

Although the zodiacal emission is main background source in the short wavelength of
far-IR range in low ecliptic latitude regions, it will not contribute to the
fluctuations on the large scales because the zodiacal light is generally smooth on
scales smaller than typical resolution of IR observations \cite{reach95,kelsall98}.
From the analysis of the \textit{ISO} data, $\acute{\rm A}$brah$\acute{\rm a}$m et
al. \shortcite{abra97} searched for the brightness fluctuations in the zodiacal
light at 25 $\mu$m with 5 fields of $\sim$~0.5$^\circ$~$\times$~0.5$^\circ$ at low,
intermediate, and high ecliptic latitudes. They found that an upper limit to the
fluctuations of 0.2 per cent of the total brightness level was estimated for an
aperture of 3$^\prime$ diameter. This amount of fluctuations would not cause any
significant noise.

Therefore, the sky confusion noise is mainly related to the spatial properties of
the cirrus. In many cases, the power spectrum of the dust emission can be expressed
as a simple power-law. Using the \textit{IRAS} data at 100 $\mu$m, Gautier et al.
\shortcite{gautier92} computed the power spectrum $P$ of the spatial fluctuations of
cirrus emission as a function of spatial frequency $k$, for angles between
4$^\prime$ and 400$^\prime$.
\begin{equation}
P = P_0 \left( \frac{k}{k_0} \right)^{\alpha} = P_0 \left( \frac{d_0}{d}
\right)^{\alpha}, \label{eqn_ps}
\end{equation}
where $d$ represents the angular scale corresponding angular frequency ($k =
\frac{2\pi}{d}$). The subscript 0 on $k$ and $d$ denotes a reference scale, $P_0$ is
the powers at $k=k_0$, and $\alpha$ is the index of the power spectrum. Since the
second order structure function is proportional to power spectrum representing the
spatial structure of cirrus, the sky confusion noise $N$ on a scale $d$
corresponding to the width of the measurement aperture scales as:
\begin{equation}
N \propto \left( \frac{d}{d_0} \right)^{1- \frac{\alpha}{2}} \cdot
P_0^{\frac{1}{2}}. \label{eqn_rel_strps}
\end{equation}
HB90 extended the work by Gautier et al. \shortcite{gautier92} at $\lambda =
100~\mu$m in order to estimate the sky confusion at all wavelengths, using the
empirical relationship, $P_0 \propto \langle I_0 \rangle^3$ and $\alpha = -3$ in
Gautier et al. \shortcite{gautier92}. They found an approximation for the cirrus
confusion noise as follows (hereafter HB90 formula):
\begin{equation}
N = \zeta  \left( \frac{\lambda}{100 ~\mu \rm m} \right)^{2.5} \left( \frac{D_t}{1
~\rm m} \right)^{-2.5}\left( \frac{\langle I_{\lambda} \rangle}{1 ~\rm MJy\,
sr^{-1}} \right)^{1.5} {\rm mJy}, \label{eqn_strn_hb}
\end{equation}
where $\zeta$ is a constant, $\lambda$ the wavelength of the measurement, $D_t$ the
diameter of the telescope, and $\langle I_{\lambda} \rangle$ is the mean brightness
at the observation wavelength. They set the constant $\zeta$ to be 0.3.

This indicates that the sky confusion depends upon both the variation of the surface
brightness in the background structure and the resolution of the telescope.
Consequently, the noise becomes less significant for larger aperture sizes.

\section{GENERATION OF CIRRUS MAP}\label{sec:gen_dmap}
In order to investigate the sky confusion for the present and upcoming infrared
space missions with a high resolution, we need the information on the behavior of
cirrus emission in very small scales. Since observationally available data have
rather low resolution, we need to add high resolution component. In this section, we
describe the method of extending the low resolution data to high resolution. For the
observational low resolution data, we used the all-sky 100 $\mu$m dust map generated
from the \textit{IRAS} and \textit{COBE} data by Schlegel, Finkbeiner, and Davis
(1998; hereafter SFD98).

\subsection{Fluctuations at Higher Spatial Resolution}

\subsubsection{Measured Power Spectrum}

\begin{figure}
    \centering \centerline{
    \psfig{figure=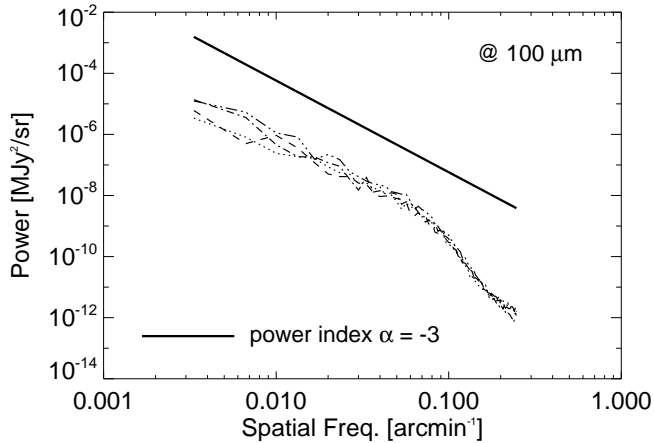,height=6.5cm}}
    \caption{Measured power spectrum of dust emission in the dust map by SFD98
    (Schlegel, Finkbeiner \& Davis 1998). The four curves represent four patches
    selected in the Northern and the Southern Galactic sky at $b = |50|^{\circ}$.}
   \label{fig_mps_SFD98}
\end{figure}

Fig. \ref{fig_mps_SFD98} shows the measured power spectrum in the dust maps of SFD98
at a Galactic latitude of $b = |50|$ degrees. These power spectra are well fitted to
power laws of index -2.9. However, the power drops at higher frequencies
corresponding to the map resolution of $\sim$ 6.1 arcmin. This breakdown of the
power spectrum is due to the large beam size of IRAS map. Although we can recover
the small-scale fluctuation by the deconvolution of a point spread function (PSF),
there is clearly some limitation. We need to generate the dust map including the
contributions from small-scale fluctuations in order to study for the planned
present and future missions with high resolution ($<$~1 arcmin). We obtain such high
resolution map by adding small-scale structure of cirrus emission to the
low-resolution map of SFD98 assuming that the small-scale fluctuations also follow
the estimated power spectrum with the same power-law index, as described above.

\subsubsection{Small Scale of Fluctuations}\label{subsec:sim_de}

The power, $P(k)$, is defined as the variance of the amplitude in the fluctuations:
\begin{equation}
P(k) \equiv \langle\mid\delta_k\mid^2 \rangle = \frac{1}{V}\int \xi(x)
\frac{\sin(kx)}{kx} 4\pi x^2 dx,
  \label{eqn_powvar}
\end{equation}
where $\delta_k$ is the perturbation field, $\langle\mid\delta_k\mid^2\rangle$ is
the variance of the fluctuation and $\xi(x)$ is the correlation function of the
brightness field. We assume that the distribution of fluctuations is approximated as
a random Gaussian process where the Fourier components $\delta_k$ have random phases
so that the statistical properties of distribution are fully described by the power
spectrum $\mid \delta_k \mid^2$ \cite{peebles80}. In this case, we can set each
fluctuation within a finite grid in the frequency domain by a random Gaussian
process of the amplitude of each fluctuation considering the realization of a volume
for the sample embedded within a larger finite volume \cite{gp90,park94,peacock99}.
We assign Fourier amplitudes randomly within the above distribution in the finite
volume and assign phases randomly between 0 and 2$\pi$. Since the field used in this
simulation is small ($<$ 10 degree), we can take the small-angle approximation and
treat the patch of sky as flat \cite{white99}. In the flat sky approximation, we
obtain the power spectrum and generate a patch of the dust map in cartesian
coordinates.

\begin{figure*}
    \centering \centerline{
    \psfig{figure=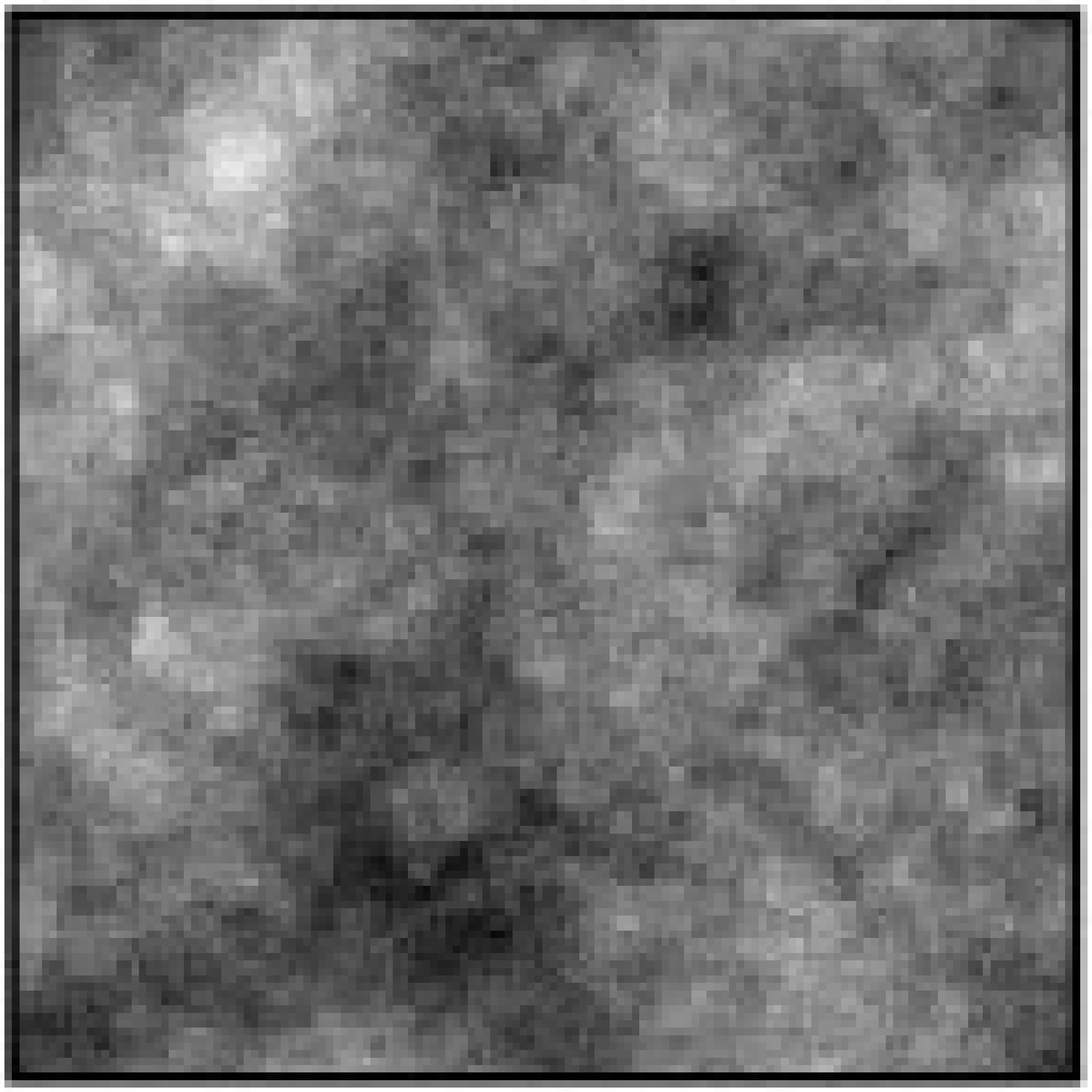, height=5cm}
    \psfig{figure=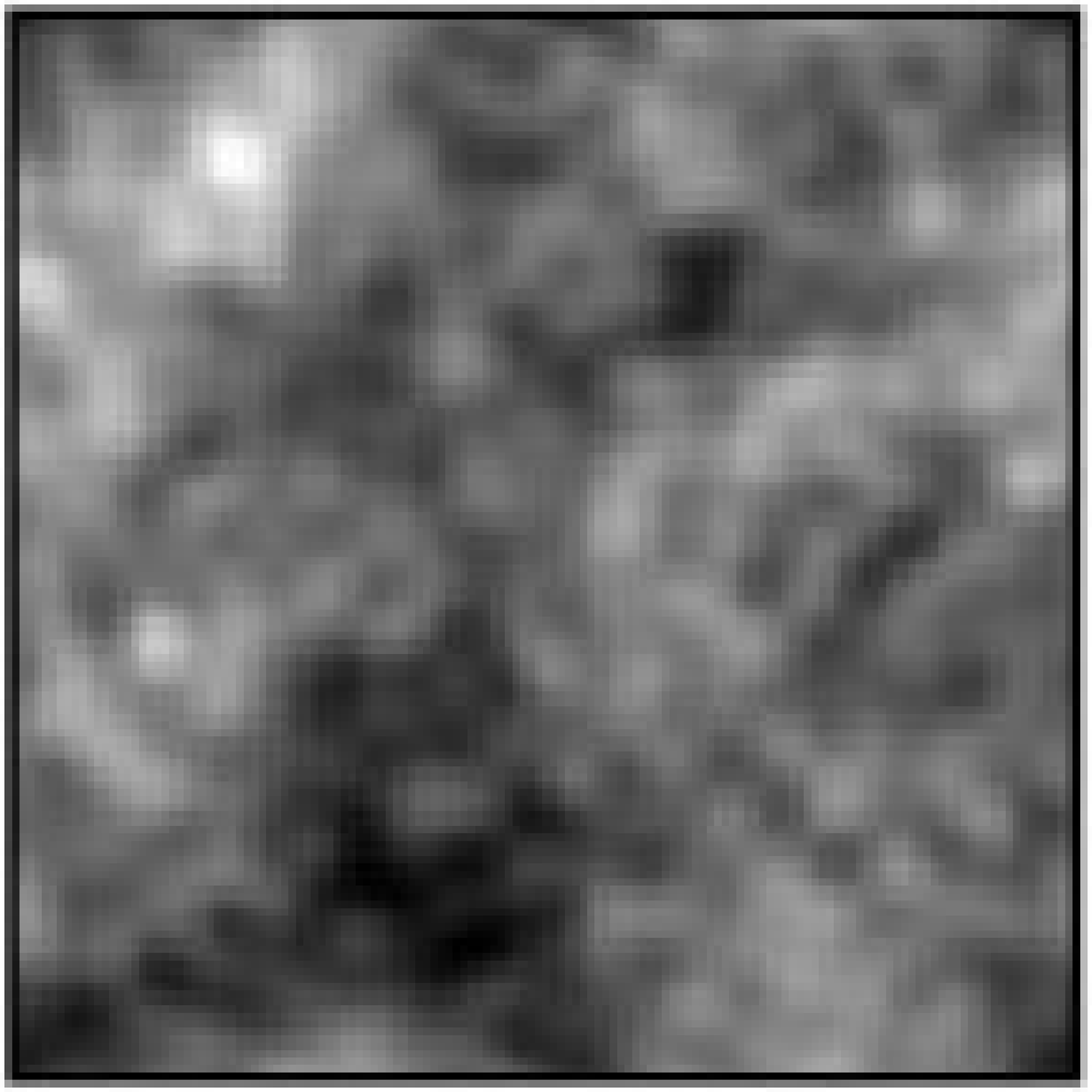, height=5cm}
    \psfig{figure=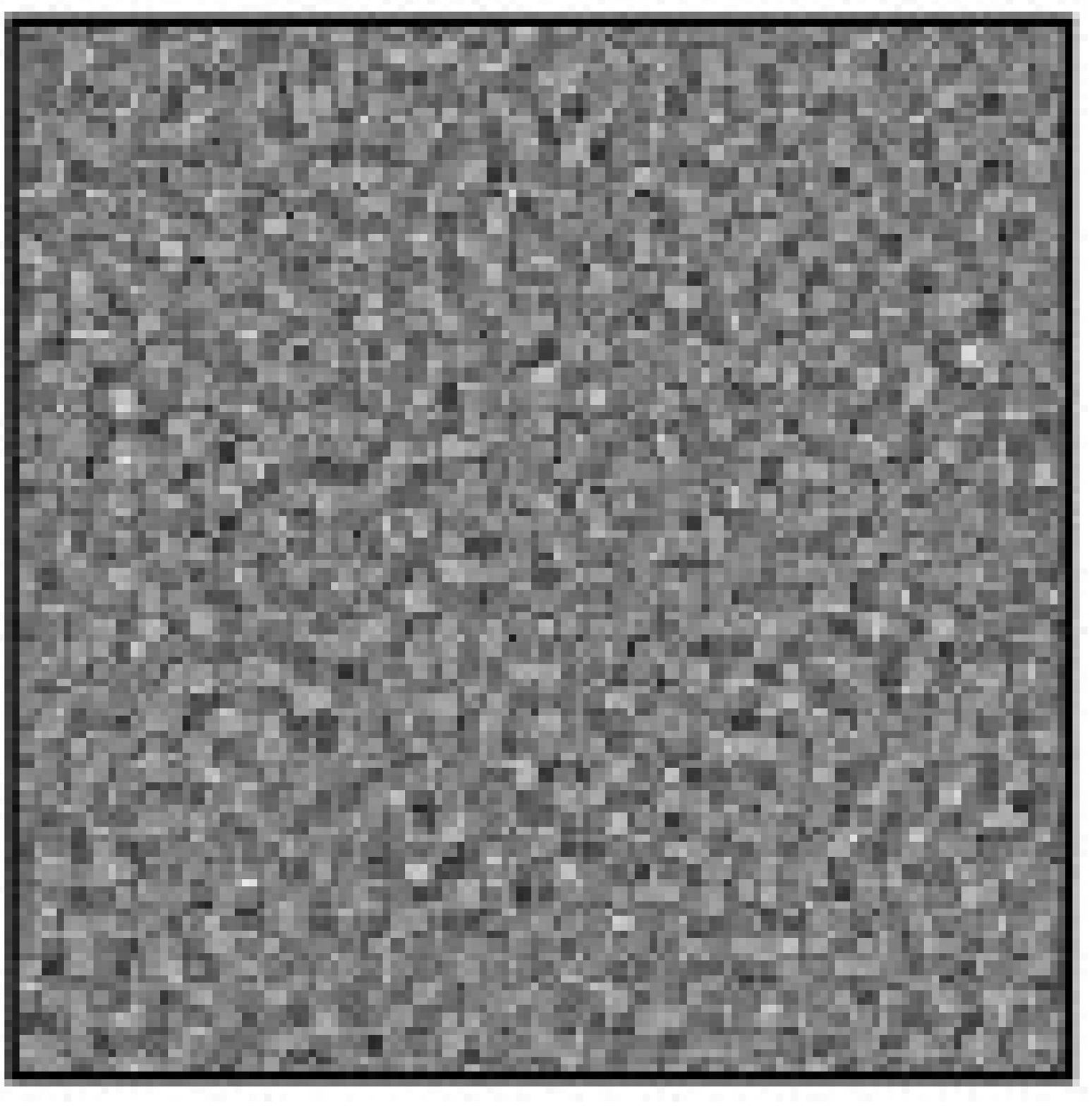, height=5cm}  }
    \centerline{
    \psfig{figure=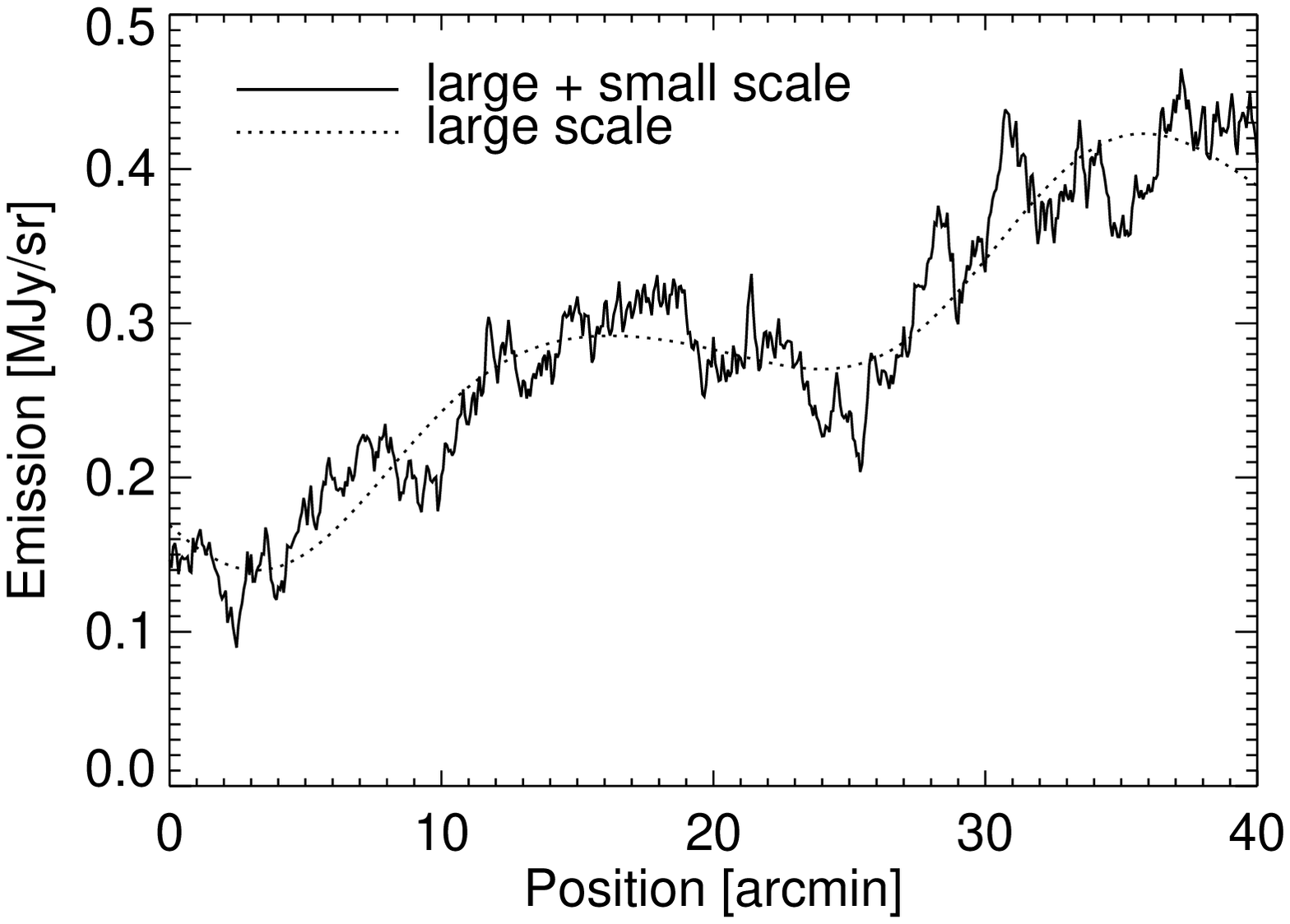, height=8cm}   }
    \caption{Simulated dust emission map (upper) and the profile of map (lower).
   The upper-left panel shows the simulated image assuming a power spectrum with
   a power index of -3. The upper-middle panel and the upper-right panel show only
   large-scale fluctuations and small-scale fluctuations, respectively. The lower
   panel shows the one-dimensional profile for a selected part of the upper-left
   and the upper-middle panel.}
   \label{fig_dmap_all}
\end{figure*}

We generate a realistic distribution of the Galactic emission in the following
manner. The basic data for the information of the large-scale structure are obtained
from the low resolution all-sky map by SFD98. We add the simulated small-scale
structure to these basic data in the Fourier domain, where the power spectrum of the
small-scale structure follows that of the large-scale structure. Fig.
\ref{fig_dmap_all} shows our simulated emission map including small-scale
fluctuations. The left panel of Fig. \ref{fig_dmap_all} shows the simulated dust
emission image corresponding to a power spectrum with $\alpha = -3$. The middle
panel includes only the emission above the resolution of the dust map by SFD98,
$\sim$ 6.1 arcmin, (large-scale emission) while the right panel shows the emission
above the resolution of the dust map by SFD98 (separated in Fourier domain, i.e.,
small-scale emission). The lower panel shows the profiles for selected areas of two
images (upper-left and upper-middle panels). We find in this simulation that the
emission including the high resolution, small-scale component (above the resolution
of the dust map by SFD98 to a resolution of 4 arcsec) reflects the trend of the
large-scale emission (above the resolution of SFD98 dust map).

\begin{figure*}
    \centering \centerline{
    \psfig{figure=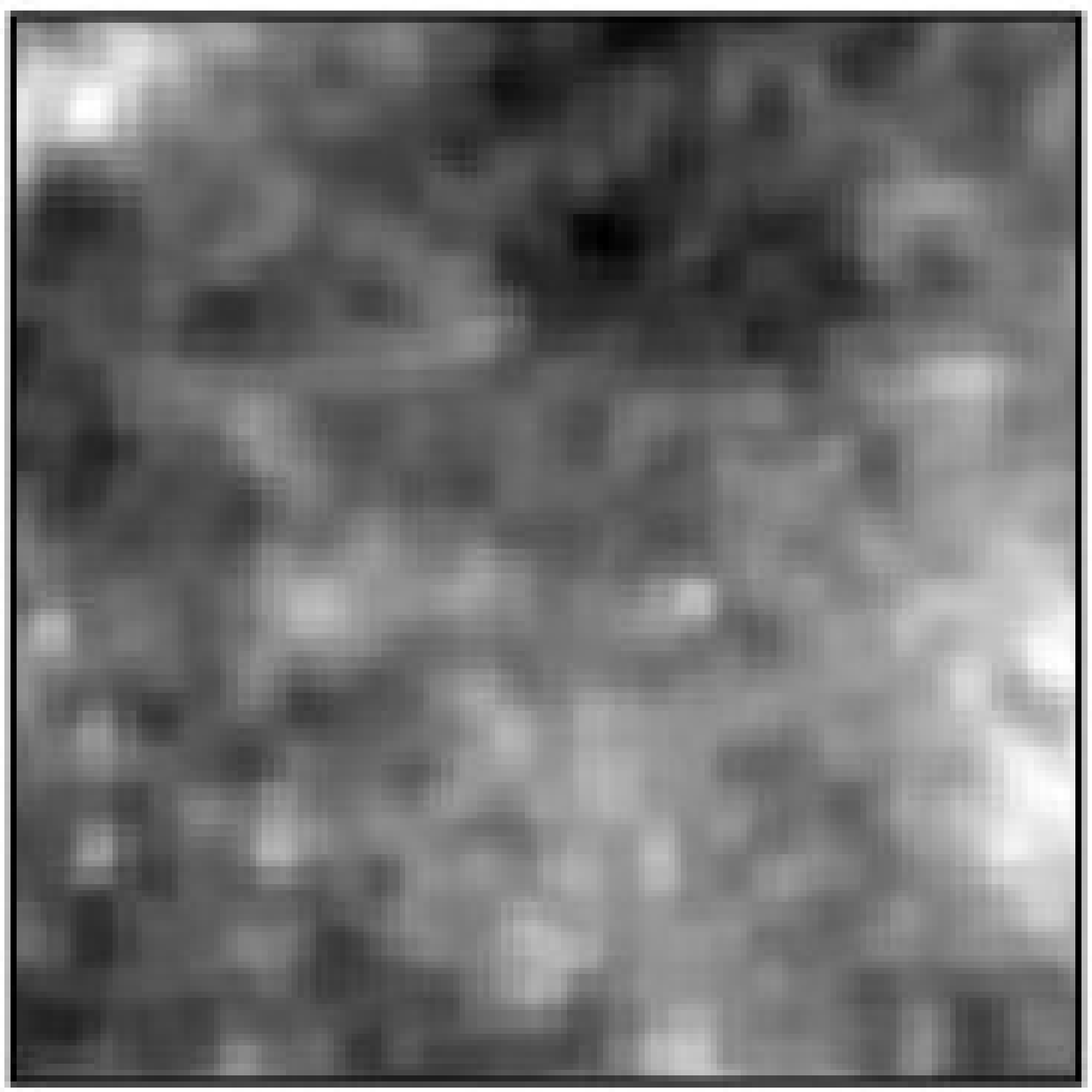, height=6cm}
    \psfig{figure=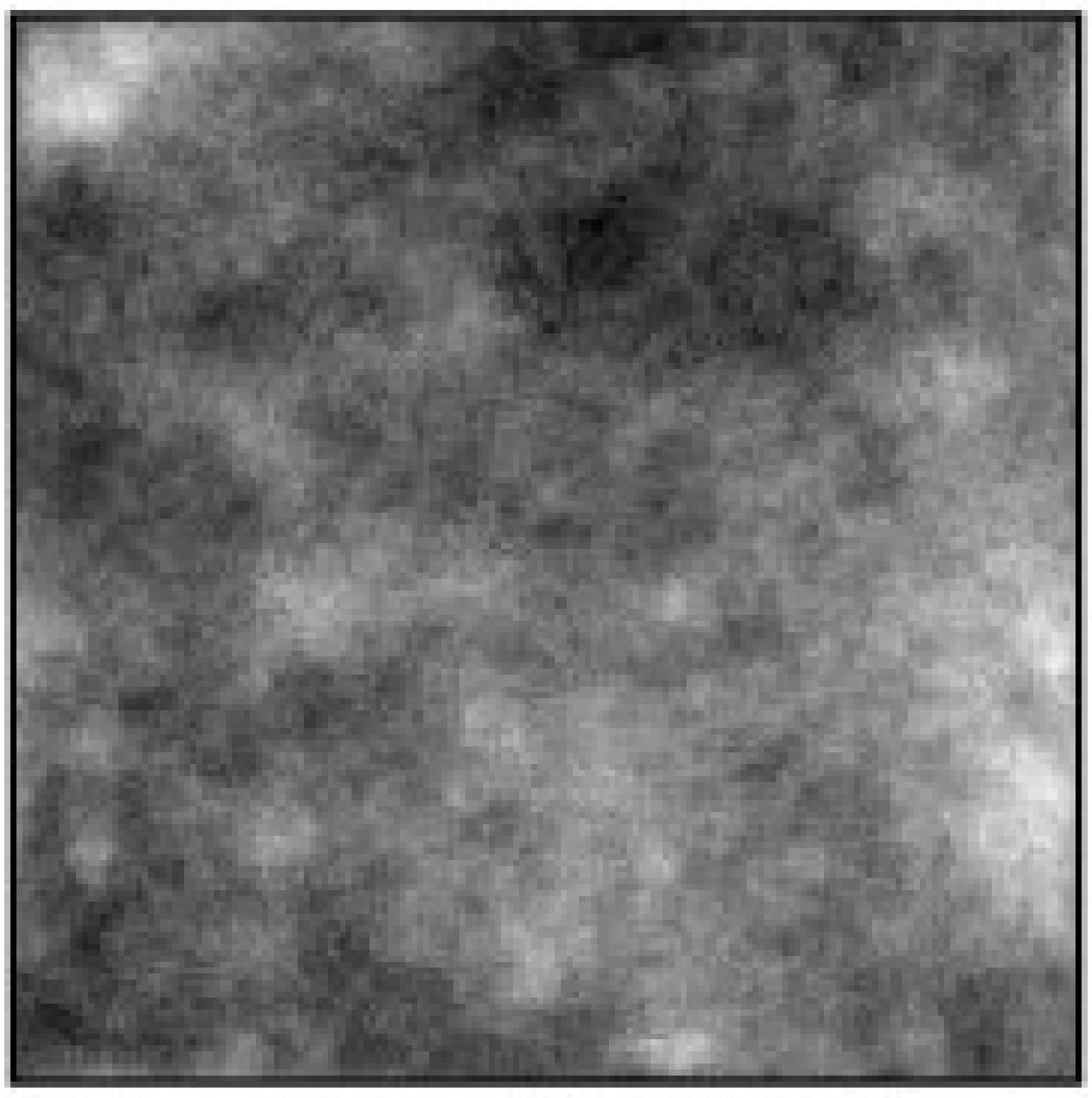, height=6cm}     }
    \centerline{
    \psfig{figure=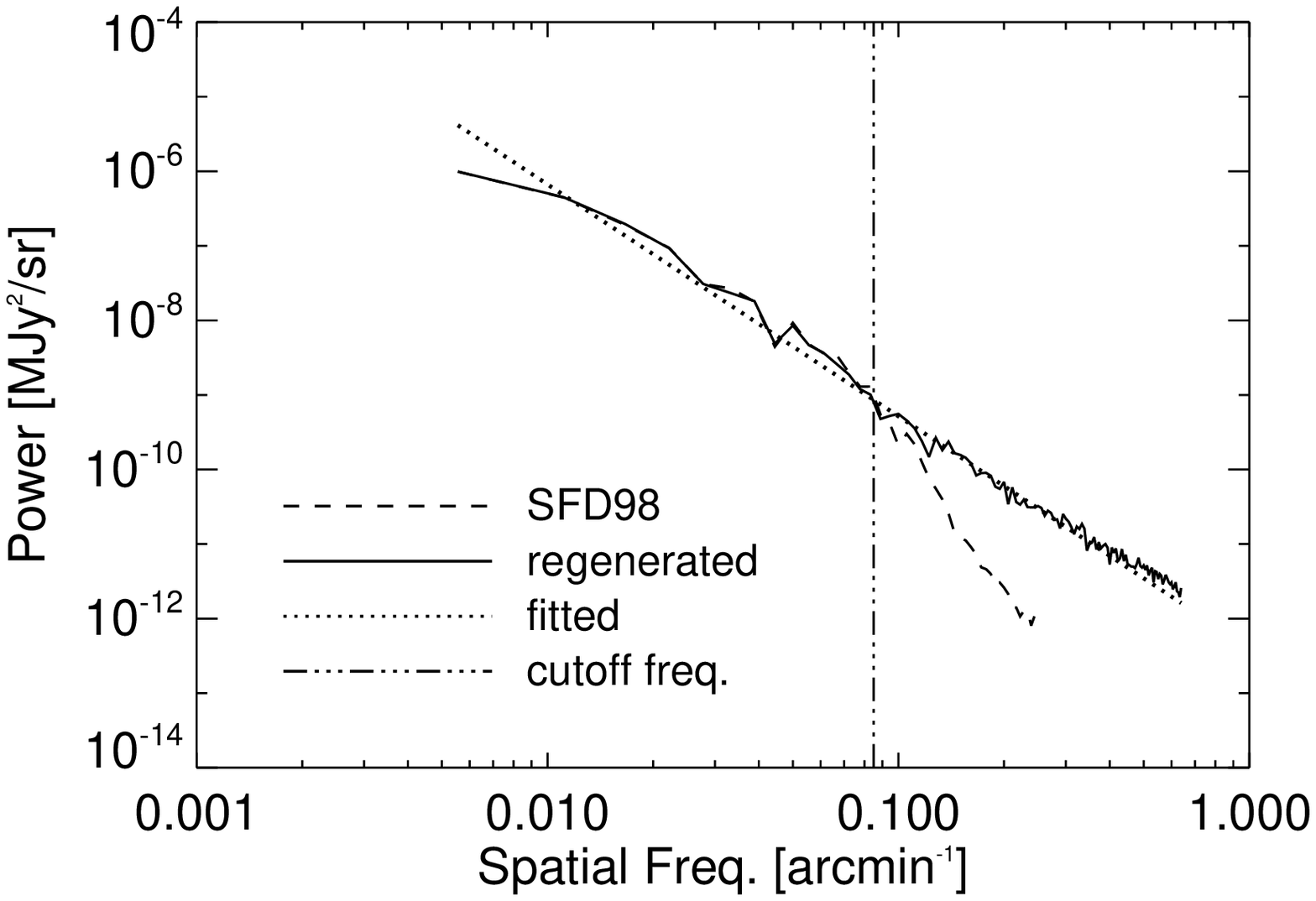, height=9cm}   }
   \caption{Patch of SFD98 dust map, regenerated patch (upper panel) and the estimated
   power spectrum (lower panel). The upper-left panel is a patch of the SFD98 dust map
   at the Galactic latitude of 50 degree and the upper-right panel is the regenerated
   patch based upon the patch from the SFD98 dust map. The dashed and
   solid lines in the lower panel show the estimated power spectrum of the upper-left
   and the upper-right panels, respectively. Note that the Nyquist frequency in the power spectrum
   of the upper-right panel is 7.5 arcmin$^{-1}$, but we only plot to
   $\sim$ 0.5 arcmin$^{-1}$. The dotted line shows the fit to the  power spectrum below
   the spatial cutoff frequency.}
   \label{fig_dmap_gb50}
\end{figure*}

We obtain a patch of the dust map including small-scale fluctuations by summing the
large-scale component of SFD98 map and the small-scale component of the simulated
emission in the Fourier domain. According to this scheme of Fourier power spectrum
analysis, the cutoff spatial frequency of the dust map by SFD98 is set to the
Nyquist limit, i.e. a half of the spatial frequency corresponding to the resolution
of the dust map by SFD98. We use the power spectrum fitted below the Nyquist
sampling limit in order to extend the power spectrum to higher spatial frequencies.
Typically, the 2D power spectrum of a SFD98 dust map patch shows the presence of a
cross along spatial frequencies of $x$ and $y$ axis if we assume that the centre in
the spatial domain is regarded as the spatial frequency 0. This cross is caused by
the Fast Fourier Transform (FFT) algorithm that makes an ``infinite pavement'' with
the image prior to computing the Fourier transform \cite{miv02}. In order to
preserve the information of the emission at the edges, we directly use the power at
the spatial frequencies of $x$ and $y$ axis, and extrapolate the power at other
spatial frequencies (above the cutoff spatial frequency) according to the estimated
power spectrum. In Fig. \ref{fig_dmap_gb50}, we show a patch of the dust map by
SFD98 at a Galactic latitude of 50 degree (upper left), a patch regenerated by
extending the power spectrum (upper right) and the estimated power spectrum (lower
panel).

\subsection{Dust Emission at Other Wavelengths}

Assuming that the spatial structure of the dust emission is independent of
wavelength, we can obtain the dust map at other wavelengths than 100 $\mu$m by
applying an appropriate model for the Spectral Energy Distribution (SED). Since the
dust particles are small ($<$ 0.25 $\mu$m) compared with far-IR wavelengths, the
opacity does not depend upon the details of the particle size distribution, but on
the nature of the emitting material itself. In the far-IR, the opacity
$\kappa_{\nu}$ generally follows a power law:
\begin{equation}
\kappa_{\nu} \propto \nu^{\beta} \label{eqn_em_law}
\end{equation}
with frequency $\nu$.

The SED may be approximated as one-component or two-component models
\cite{schl98,fink99}. The dust temperature map is constructed from the COBE Diffuse
Infrared Background Experiment (\textit{DIRBE}) 100 $\mu$m and 240 $\mu$m data
\cite{bogg92} which was designed to search for the cosmic IR background radiation.
For a one-component moedel, the emission $I_\nu$ at frequency $\nu$ can be expressed
as
\begin{equation}
I_\nu = K_{100}^{-1}(\beta, T)\, I_{100} \, \frac{\nu^{\beta} B_{\nu}
(T)}{\nu_0^{\beta} B_{\nu_0} (T)}, \label{eqn_one_em_nu}
\end{equation}
where $B_{\nu}(T)$ is the Planck function at temperature $T$, $I_{100}$ is the
\textit{DIRBE}-calibrated 100 $\mu$m map, $K_{100}^{-1}(\beta, T)$ is the colour
correction factor for the \textit{DIRBE} 100 $\mu$m filter when observing a
$\nu^{\beta} B_{\nu}(T)$ spectrum (\textit{DIRBE} Explanatory Supplement 1995).
Although the generated temperature maps have relatively low resolution (1.3$^\circ$)
compared with our simulated dust map patch, we interpolate this map to small grid
sizes ($<$ 10 arcsec). Taking the emissivity model with $\beta = 2$ \cite{dl84}, we
can obtain the dust temperature from the \textit{DIRBE} 100 $\mu$m/240 $\mu$m
emission ratio.

\begin{figure}
    \centering \centerline{
    \psfig{figure=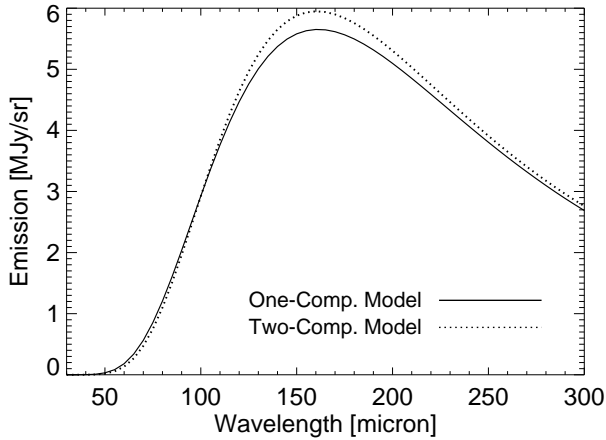,height=6.5cm}}
   \caption{Comparison between the one-component dust model and the two-component
   dust model for one small patch. The dust emission of the two-component model
   in the wavelength range from 120 $\mu$m to 200 $\mu$m is slightly higher than
   that of the one-component model due to the dominant contribution by carbon grains.}
   \label{fig_dust_model}
\end{figure}

Based upon laboratory measurements, a multicomponent model for interstellar dust has
been constructed by Pollack et al. \shortcite{poll94}. In order to solve the
inconsistency of the $\nu^2$ emissivity model in the 100 $-$ 2100 GHz (3000 $-$ 143
$\mu$m) emission, Finkbeiner et al. \shortcite{fink99} used a two-component model
where diverse grain species dominate the emission at different frequencies in order
to fit the data of the COBE Far Infrared Absolute Spectrophotometer (FIRAS).
Assuming that each component of the dust has a power-law emissivity over the FIRAS
range, Finkbeiner et al. \shortcite{fink99} constructed the emission $I_\nu$ in
multicomponent model:
\begin{equation}
I_\nu = \frac{\sum_i ~f_i ~Q_i(\nu) ~B_{\nu}(T_i)}{\sum_i ~f_i ~Q_i(\nu_0)
~B_{\nu_0}(T_i) ~K_{100}(\beta_i, T_i)} \, I_{100}, \label{eqn_mul_em_nu}
\end{equation}
where $f_i$ is a normalization factor for the $i$-th grain component, $T_i$ is the
temperature of component $i$, $K_{100}$ is the \textit{DIRBE} colour-correction
factor and $I_{100}$ is the SFD98 100 $\mu$m flux in the \textit{DIRBE} filter. The
emission efficiency $Q_i(\nu)$ is the ratio of the emission cross section to the
geometrical cross section of the grain component $i$. In order to obtain the
temperature of each component, we further need effective absorption opacity defined
by
\begin{equation}
\kappa_i^* = {\int_0^\infty \kappa_i^{\rm abs} J_{\rm ISRF} (\nu) d\nu\over
\int_0^\infty J_{\rm ISRF} (\nu) d\nu},
\end{equation}
where $\kappa_i^{\rm abs}$ is the absorption opacity of the $i$-th component, and
$J_{\rm ISRF}$ is the mean intensity of interstellar radiation field. Finkbeiner et
al. \shortcite{fink99} assumed that the normalization factors do not vary with
locations and size independent optical properties of dust grains. The emission
efficiency factor $Q_i$ at far-IR is further assumed to follow a power-law with
different indices ($\beta$) for different dust species. In the present work, we
adopted the `best-fitting' two-component model by Finkbeiner et al.
\shortcite{fink99}: $\beta_1 = 1.67$, $\beta_2$=2.70, $f_1=0.0363$, $f_2=0.9637$,
and $q_1/q_2 =13.0$, where $q_i=\kappa_i^{\rm abs}(\nu_0)/\kappa_i^*$ which
represents the ratio of far-IR emission cross section to the UV/optical absorption
cross section. The reference frequency $\nu_0$ is that corresponding to wavelength
100 $\mu$m.

If we further assume that the interstellar radiation field has constant spectrum,
the temperature of each component can be uniquely determined by the far-IR spectrum
represented by the \textit{DIRBE} 100 $\mu$m/240 $\mu$m ratio. A two-component model
provides a fit to an accuracy of $\sim$ 15 per cent to all the FIRAS data over the
entire high-latitude sky. In Fig. \ref{fig_dust_model}, we see the dust emission for
the one-component and two-component dust models [see Schlegel et al.
\shortcite{schl98}; Finkbeiner et al. \shortcite{fink99}]. The two-component model
agrees better with the FIRAS data in the wavelength range longer than 100 $\mu$m
where the dust emission estimated from one-component model is significantly lower
than the estimate from the two-component model.

In two models, the contribution of the small grains resulting in an excess below 100
$\mu$m is not considered. Since there is no significant difference between models
below 100 $\mu$m while the dust emission of the two-component model is slightly
higher than that of the one-component model in wavelengths ranging from 120 to 200
$\mu$m, we use the two-component model in our calculations.

Through a PSF convolution at each wavelength and a wavelength integration over a 5
$\mu$m wavelength grid, we obtain the high resolution dust map in other bands.

\section{FLUCTUATION ANALYSIS FOR SKY CONFUSION NOISE}\label{sec:stat_analy_scn}

Among the parameters affecting the sky confusion noise, most of them depend upon the
mean brightness, the spatial structure of the cirrus, and the observing wavelength,
as seen in equation (\ref{eqn_strn_hb}). In Table \ref{tab_inst_para}, we list the
basic instrumental parameters of present and future IR space missions; the aperture
of the telescope, Full Width at Half Maximum (FWHM) of the beam profile and the
pixel size for each detector. For comparison with previous studies
\cite{herb98,kiss01}, we include the specifications for \textit{ISO}. We select a
short wavelength band (SW) and a long wavelength band (LW) for each mission.

\begin {table*}
\centering
\caption {Instrumental parameters for various space missions.} \label{tab_inst_para}
\vspace{5pt}
\begin{tabular}{@{}cccccccc}
\hline\vspace{-5pt} \\
& Aperture & Wavelength \span\omit & FWHM $^a$\span\omit & Pixel size\span\omit \vspace{5pt} \\
 & (meter) & ($\mu$m)\span\omit & (arcsec)\span\omit & (arcsec)\span\omit \vspace{5pt} \\
Space Mission & & SW & LW & SW & LW & SW & LW \vspace{5pt}
\\\hline \vspace{-10pt}
\\ \textit{ISO} $^b$ & 0.6 & 90 & 170 & 31.8 & 60 & 46 & 92 \vspace{5pt}
\\ \textit{Spitzer} $^c$ & 0.85 & 70 & 160 & 16.7 & 35.2 & 9.84 & 16 \vspace{5pt}
\\ \textit{ASTRO-F} $^d$ & 0.67 & 75 & 140 & 23 & 44 & 26.8 & 44.2 \vspace{5pt}
\\ \textit{Herschel} $^e$ & 3.5 & 70 & 160 &  4.3 & 9.7 & 3.2 & 6.4 \vspace{5pt}
\\ \textit{SPICA} & 3.5 & 70 & 160 &  4.3 & 9.7 & 1.8 & 3.6 \vspace{5pt}
\\ \hline
\end{tabular}

\medskip
\begin{flushleft}
{\em $^a$} FWHM of diffraction pattern. \\
{\em $^b$} Two ISOPHOT filters (C1\_90 in SW band and C2\_170 in LW band). \\
{\em $^c$} MIPS bands for the \textit{Spitzer} mission. \\
{\em $^d$} \textit{ASTRO-F/FIS} (Far Infrared Surveyor) has a WIDE-S band in SW and
WIDE-L band in LW. \\
{\em $^e$} PACS have `blue' array in short wavelength (60-85$\mu$m or 85-130$\mu$m)
and the `red' array in long wavelength (130-210$\mu$m).
\end{flushleft}
\end{table*}

In order to examine the dependency of the sky confusion noise on the instrumental
parameters, we list sky confusion $N$ estimated from HB90 formula for each mission
considered in this work in Table \ref{tab_const_HB}. As the aperture of the
telescope becomes larger or the wavelength becomes shorter, sky confusion $N$ should
become correspondingly smaller. In Section \ref{sec:gen_dmap}, we obtained the dust
maps extended to high spatial resolution over a wide spectral range. With this
simulated dust map, we estimate the sky confusion noise for various space mission
projects.

\begin {table}
\centering
\caption {Sky confusion noise estimated from HB90 formula for each space mission.
The instrumental parameters for each mission are given in Table \ref{tab_inst_para}.
The mean brightness here is fixed to be 1 MJy~sr$^{-1}$.} \label{tab_const_HB}
\vspace{5pt}
\begin{tabular}{@{}ccc}
\hline\vspace{-5pt} \\
& $N$ (mJy) \span\omit \vspace{5pt} \\
Space Mission & SW & LW \vspace{5pt}
\\\hline \vspace{-10pt}
\\ \textit{ISO} & 0.83 & 4.05 \vspace{5pt}
\\ \textit{Spitzer} & 0.18 & 1.46 \vspace{5pt}
\\ \textit{ASTRO-F} & 0.40 & 1.89 \vspace{5pt}
\\ \textit{Herschel}  & 0.0054 & 0.042 \vspace{5pt}
\\ \textit{SPICA} & 0.0054 & 0.042 \vspace{5pt}
\\ \hline
\end{tabular}
\end{table}

\subsection{Selected Regions}

\begin{figure}
    \centering \centerline{
    \psfig{figure=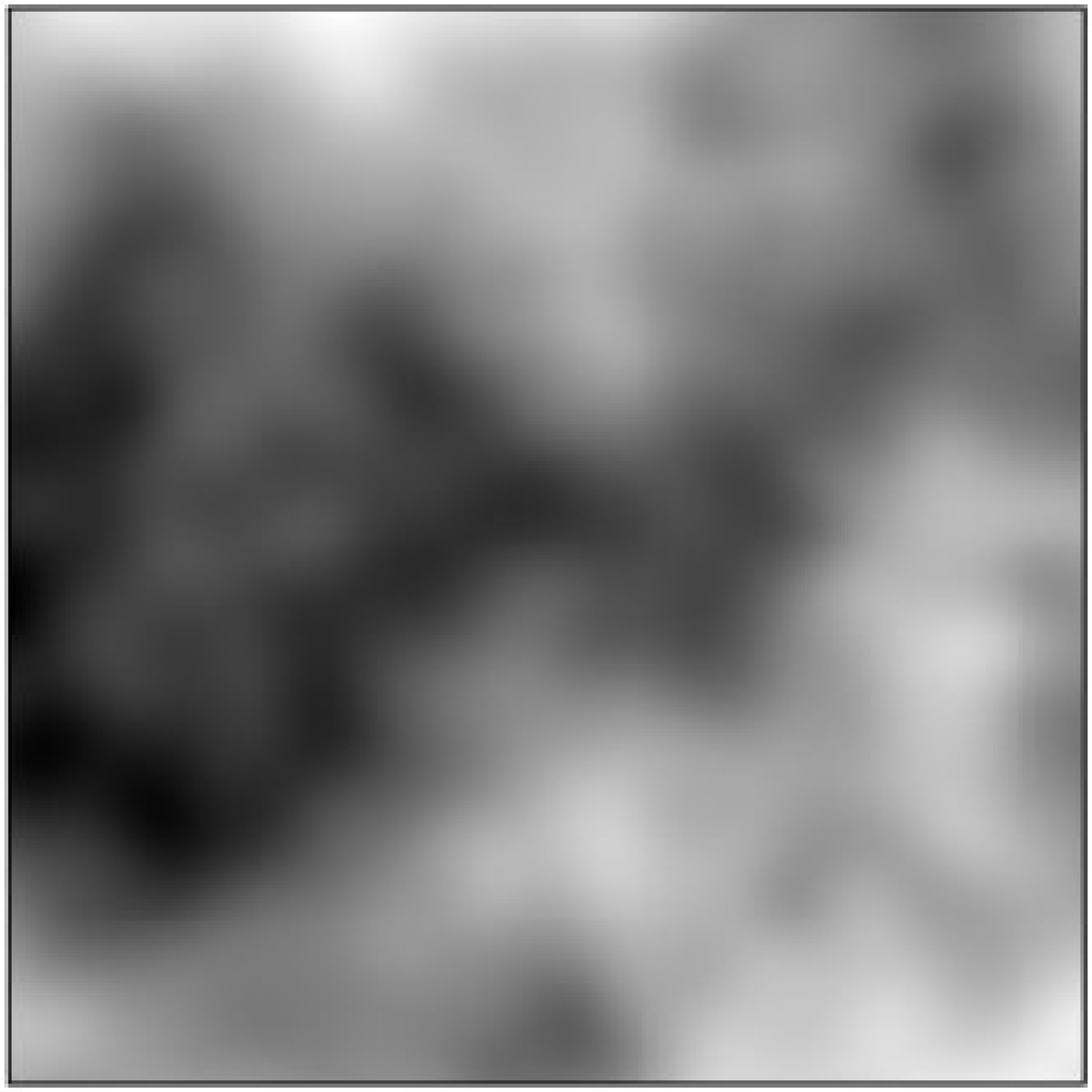, height=3.5cm}
    \psfig{figure=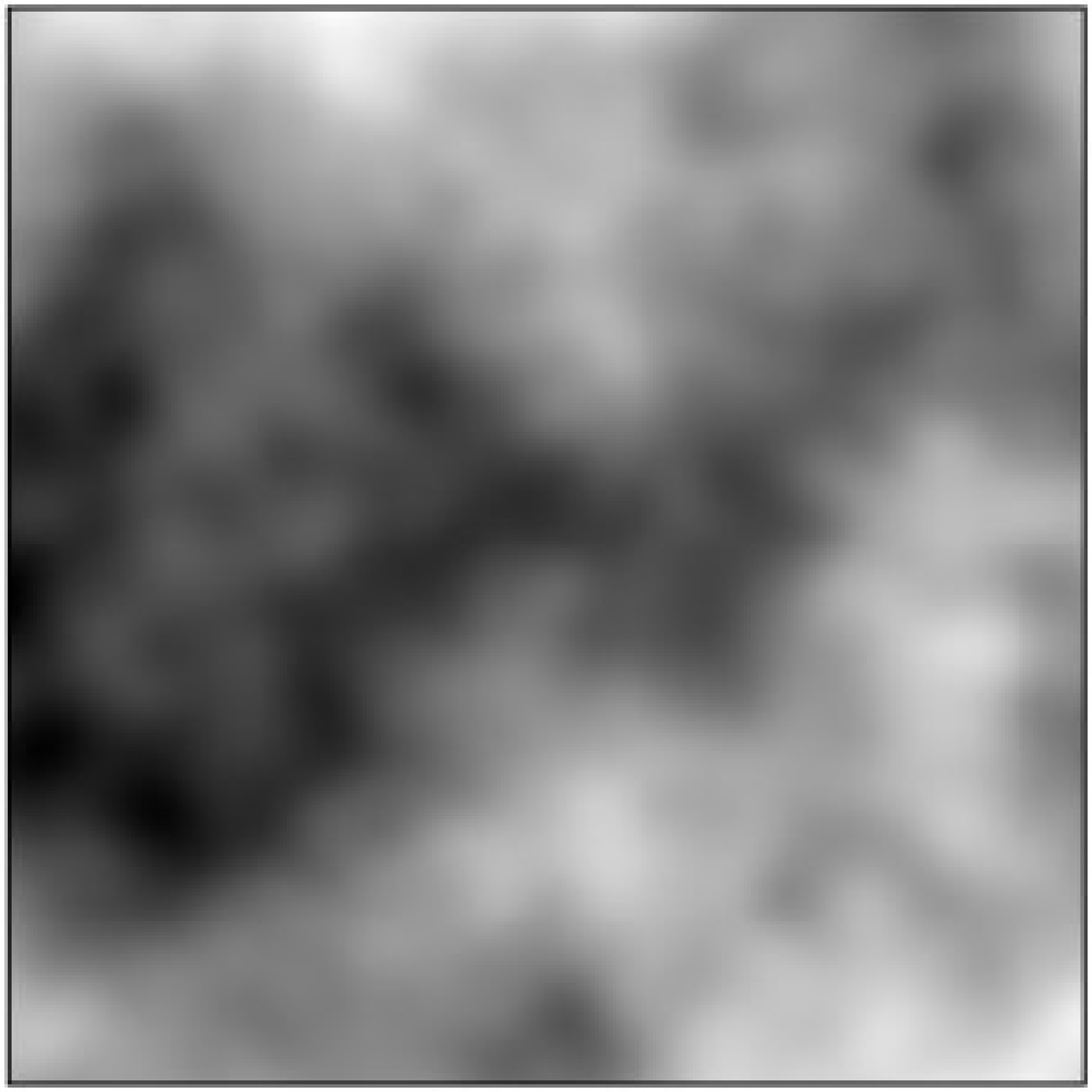, height=3.5cm} }
    \centerline{
    \psfig{figure=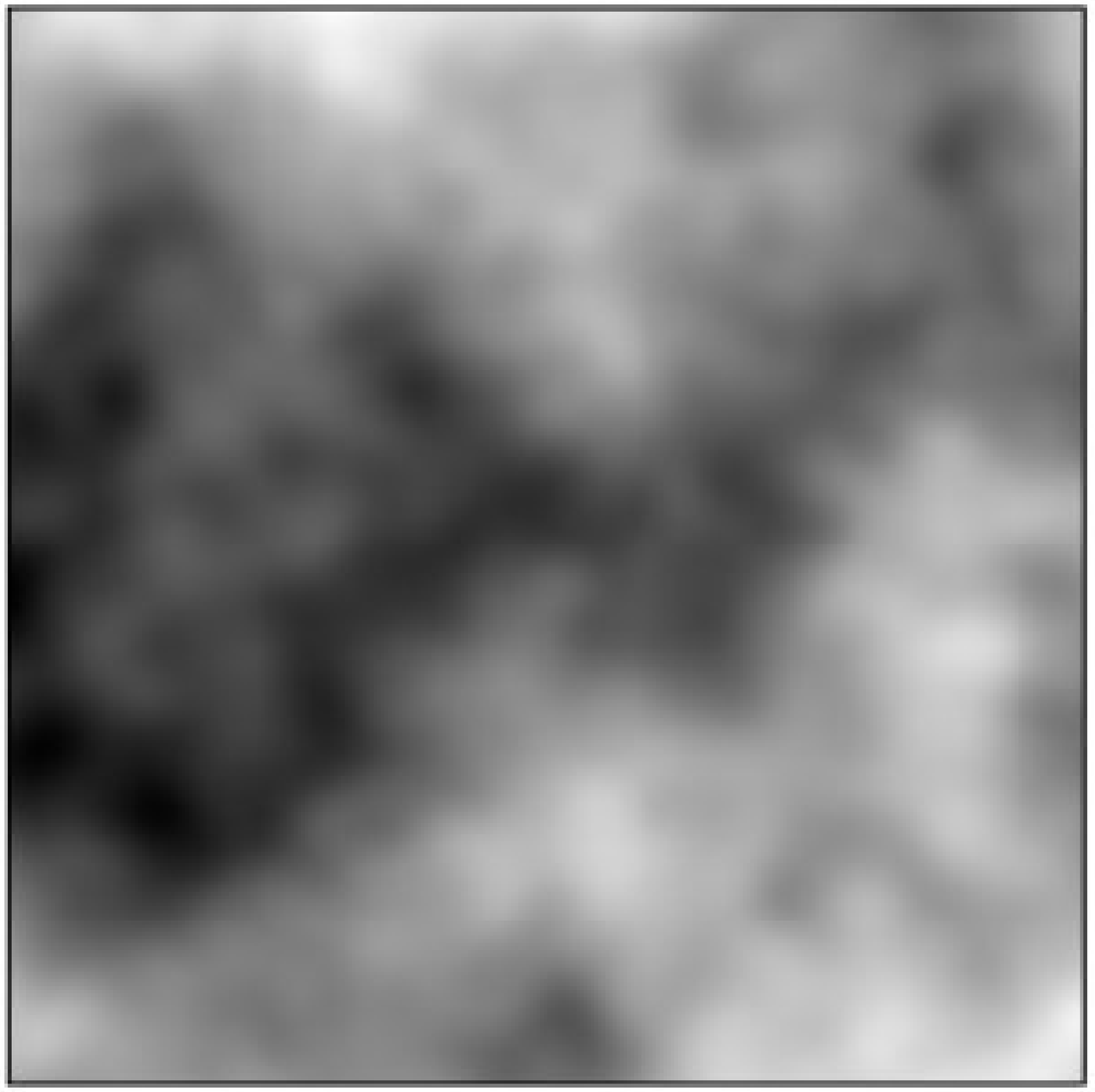, height=3.5cm}
    \psfig{figure=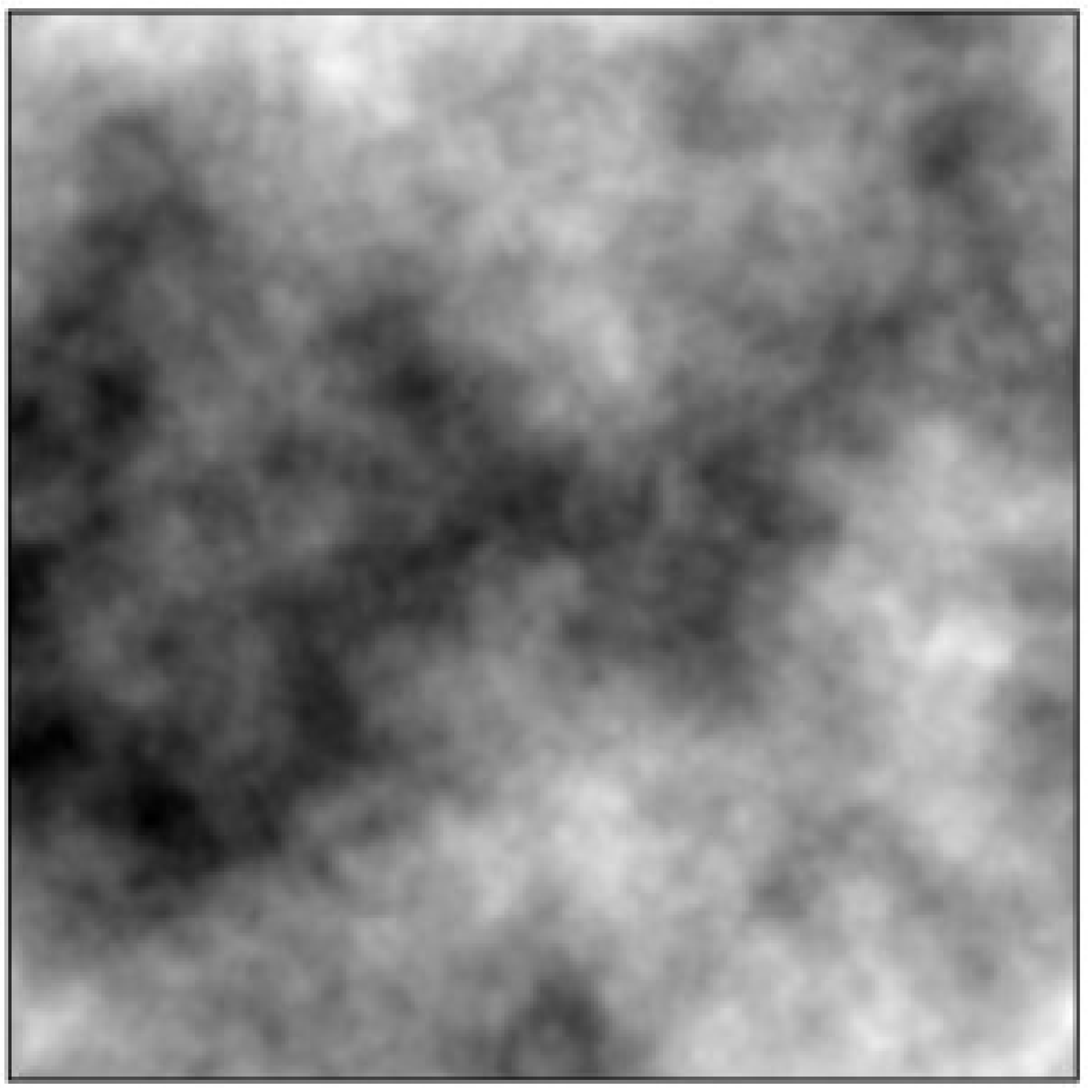, height=3.5cm}  }
   \caption{PSF-convolved patch of the dust map for space mission; \textit{ISO} (upper-left),
   \textit{ASTRO-F} (upper-right), \textit{Spitzer} (lower-left), \textit{Herschel/SPICA} (lower-right) missions.}
   \label{fig_psfcv_dmap}
\end{figure}

\begin {table}
\begin{minipage}{80mm}
\centering \caption {Properties of the selected regions. The Galactic longitude of
all patches is 0$^\circ$. I$_0$ is a mean sky brightness, $\alpha$ is the power
index of the power spectrum, and P$_0$ is the power estimated at 0.01 arcmin$^{-1}$
and 100 $\mu$m.} \vspace{10pt} \label{tab_prop_patch}
\begin{tabular}{@{}cccccc}
\hline \vspace{-10pt} \\
 & I$_0$ $^a$\span\omit \span\omit & $\alpha$ $^b$ & $\log$ P$_0$ $^c$ \vspace{5pt} \\
 & (MJy~sr$^{-1}$)\span\omit \span\omit & & (Jy$^2$~sr$^{-1}$) \vspace{5pt} \\
Region~$^a$ & 70$\mu$m & 100$\mu$m & 160$\mu$m & & \vspace{5pt}
\\\hline \vspace{-10pt}
\\ $b$=10$^{\circ}$ & 5.4 & 24.4 & 53.9 & -3.45$\pm$0.11 & 9.00$\pm$0.17 \vspace{5pt}
\\ $b$=17$^{\circ}$ & 3.5 & 18.6 & 45.3 & -3.50$\pm$0.16 & 9.05$\pm$0.24 \vspace{5pt}
\\ $b$=22$^{\circ}$ & 3.5 & 15.3 & 34.1 & -3.54$\pm$0.15 & 8.48$\pm$0.22 \vspace{5pt}
\\ $b$=28$^{\circ}$ & 2.2 & 8.9 & 24.7 & -3.50$\pm$0.15 & 7.74$\pm$0.21 \vspace{5pt}
\\ $b$=36$^{\circ}$ & 1.2 & 6.0 & 14.4 & -3.80$\pm$0.10 & 7.41$\pm$0.15 \vspace{5pt}
\\ $b$=45$^{\circ}$ & 0.6 & 2.8 & 6.2 & -3.13$\pm$0.12 & 6.39$\pm$0.18 \vspace{5pt}
\\ $b$=59$^{\circ}$ & 0.3 & 1.4 & 2.9 & -2.99$\pm$0.09 & 6.00$\pm$0.13 \vspace{5pt}
\\ $b$=70$^{\circ}$ & 0.2 & 1.2 & 2.6 & -3.20$\pm$0.10 & 6.27$\pm$0.15 \vspace{5pt}
\\ $b$=84$^{\circ}$ & 0.1 & 0.8 & 1.8 & -2.87$\pm$0.09 & 5.77$\pm$0.14 \vspace{5pt}
\\ $b$=90$^{\circ}$ & 0.1 & 0.5 & 1.4 & -2.87$\pm$0.08 & 5.66$\pm$0.12 \vspace{5pt}
\\ \hline
\end{tabular}

\end{minipage}
\end{table}

We generate the PSF-convolved patches of a dust map as a function of increasing
Galactic latitude (decreasing sky brightness) from 0.3 MJy~sr$^{-1}$ to 25
MJy~sr$^{-1}$ at 100 $\mu$m at a resolution of 1 arcsec by using the method
explained in Section \ref{sec:gen_dmap}. The size of the simulated image is
1.3$^\circ$ $\times$ 1.3$^\circ$. For the PSF, we used an ideal circular aperture
Airy pattern corresponding to the aperture size of telescopes. In Fig.
\ref{fig_psfcv_dmap}, we can see the PSF-convolved small patch of dust map
(900$\arcsec$ $\times$ 900$\arcsec$) for each space mission. As the aperture of the
telescope becomes larger, the structures that can be visible become smaller. Since
the cirrus emission generally depends upon the Galactic latitude, we select the
patches as a function of the Galactic latitude. We list the properties of selected
regions at a Galactic longitude of 0$^\circ$ among 50 patches in Table
\ref{tab_prop_patch}. The estimated power spectrum in Table \ref{tab_prop_patch}
differs from patch to patch. In order to reflect the large structure of the dust map
and reduce the discrepancies of the power spectrum between adjacent patches, we use
a large area around the patch ($\sim$ 2.5$^\circ$ $\times$ 2.5$^\circ$) in the
measurement of the power spectrum.

\subsection{Estimation of Sky Confusion Noise}

\subsubsection{Contribution of Instrumental Noise}\label{subsec:inst_noise}

In order to estimate the sky confusion noise, the structure function for the cirrus
emission patch obtained by measuring the sky brightness fluctuations is widely used
\cite{gautier92,herb98,kiss01}. The size of the measuring aperture is set to be the
FWHM of each beam profile if the detector pixel size is smaller than the FWHM of a
beam profile. Since the sky confusion noise and the instrumental noise are
statistically independent \cite{herb98,kiss01}, the measured noise $N_{\rm meas}$ is
\begin{equation}
N_{\rm meas}^2 = N^2 + \eta \cdot \sigma_{\rm inst}^2, \label{eqn_stat_strno}
\end{equation}
where $N$ is the sky confusion noise corresponding 1$\sigma$, $\sigma_{\rm inst}$ is
the instrumental noise, and $\eta$ is the contribution factor from the instrumental
noise. The contribution factor $\eta$ can be determined by the size of the
measurement aperture and the separation (see equation \ref{eqn_strno} and Fig.
\ref{fig_ref_aper}).

\subsubsection{Comparison with Other Results}

\begin{figure}
    \centering \centerline{
    \psfig{figure=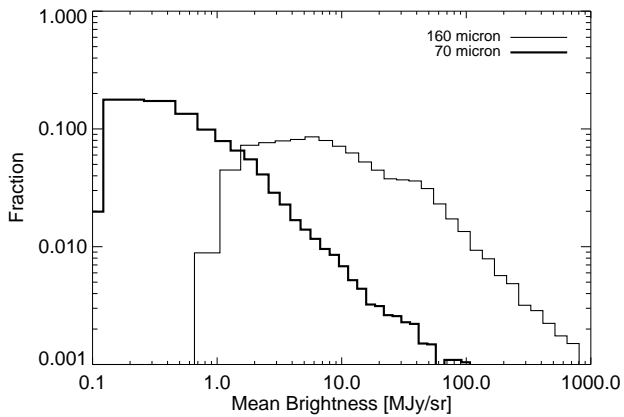, height=6cm}    }
   \caption{The fraction of the sky brightness for all sky. Note that most of the
   sky have the sky brightness below 1 MJy~sr$^{-1}$ (SW) and 15 MJy~sr$^{-1}$
   (LW). The contribution in the highest mean brightness resulted from near the
   Galactic center.}
   \label{fig_frac_em}
\end{figure}

We estimate the sky confusion noise from the patches of the simulated sky map. In
Fig. \ref{fig_frac_em}, we plot the fractional area as a function of sky brightness
over the whole sky to visualize the sky brightness distribution.
Since we consider the sky confusion caused solely by the emission from cirrus
structures, we do not include any contribution from the instrumental noise.

In order to determine a dependency of the sky confusion noise on separation, we
performed a ``calculation'' for the estimation of sky confusion noise for given mean
brightness of the sky patch for each space mission (\textit{ISO}, \textit{Spitzer},
\textit{ASTRO-F}, and \textit{Herschel/SPICA}) by systematically varying the value
of $s$ from 2 to 7, using equation (\ref{eqn_strno}), where $s$ parameter is related
to the separation $\theta$ = $sD$. Generally, larger separation causes larger sky
confusion noise because we may be estimating the fluctuations from different
structures. In practical photometry, large separations are generally used, i.e.,
$\theta$ = $sD$, $s>2$ in the configuration of Fig. \ref{fig_ref_aper}
\cite{kiss01,laur03}. As a reference, we take the estimate of the sky confusion
noise with $s = 2.5$ for a comparison of the measured sky confusion with the
photometric results given in Section \ref{sec:scn_phot}. In the source detection,
the background estimation parameter have the same role with the separation
parameter. We found the optimal value for the background estimation parameter
through the photometry (see Section \ref{sec:sub_source_det} for detailed
explanation).

\begin{figure*}
    \centering \centerline{
    \psfig{figure=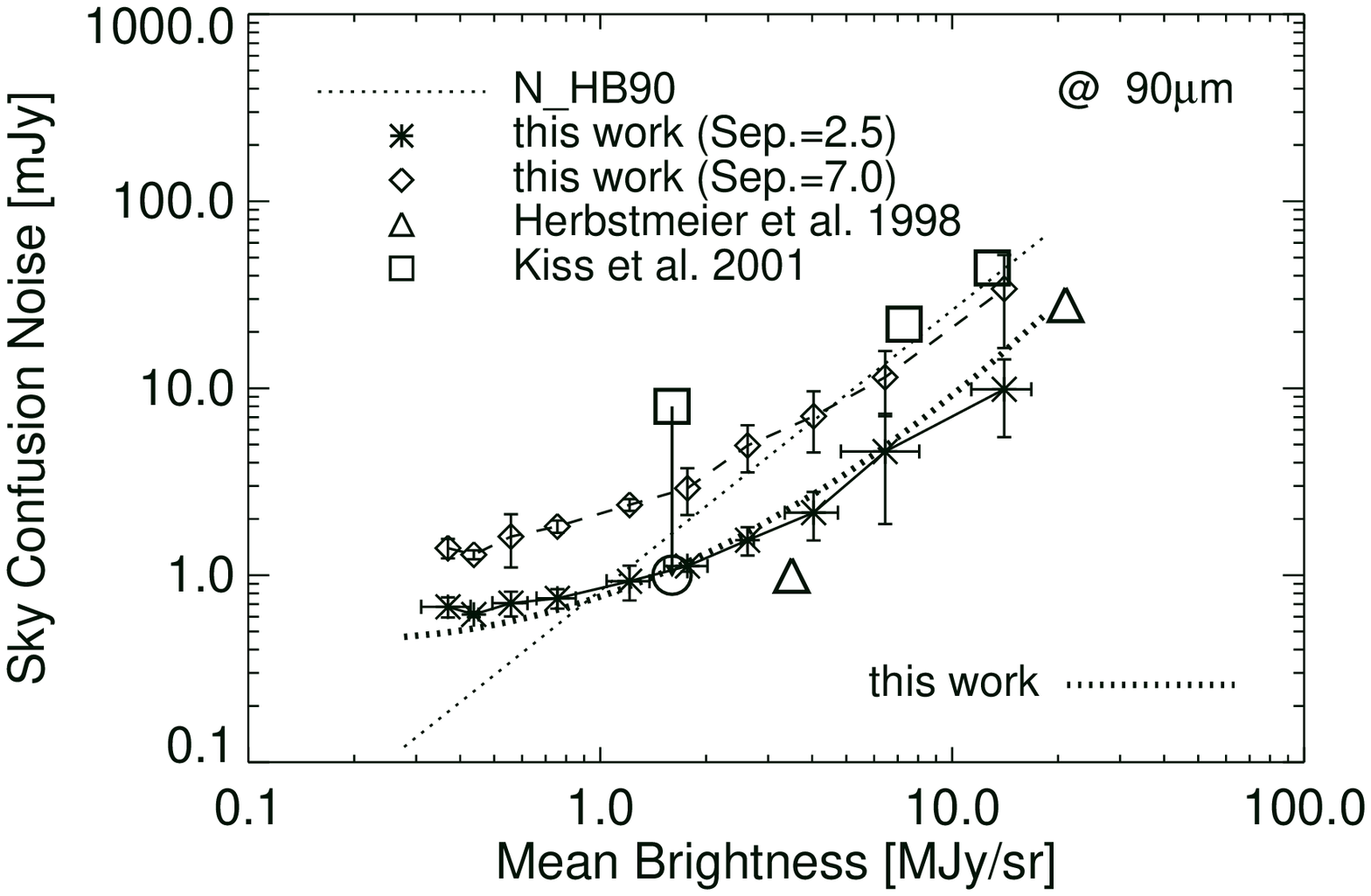, height=6.4cm}
    \psfig{figure=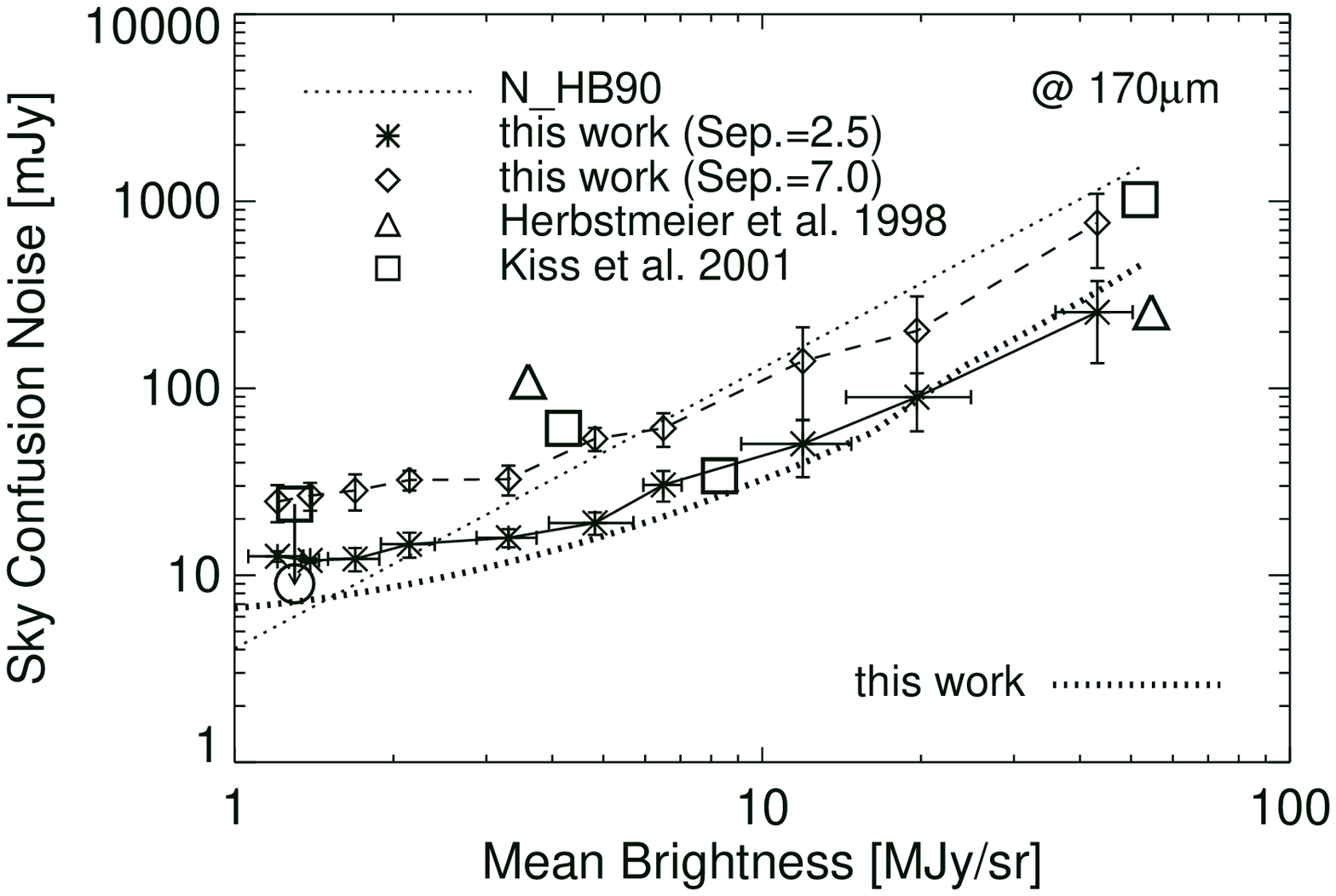, height=6.4cm}     }
    \caption{Estimated sky confusion noise for the \textit{ISO} mission. Upper and lower panels show
    the sky confusion noise at 90 $\mu$m and 170 $\mu$m, respectively. The dotted line
    shows the sky confusion noise by HB90 (Helou \& Beichman 1990). The symbols are
    the estimated sky confusion noise on averaging 5 patches with similar mean brightness.
    For comparison, we plot the estimated sky confusion noise for the larger separation of
    $s=7$. The circle symbol means the sky confusion noise correcting the
    contribution from the CFIRB. The thick dotted line is the HB90 formula to which our
    empirical relation is applied.}
   \label{fig_strn_iso}
\end{figure*}

\begin{figure*}
    \centering \centerline{
    \psfig{figure=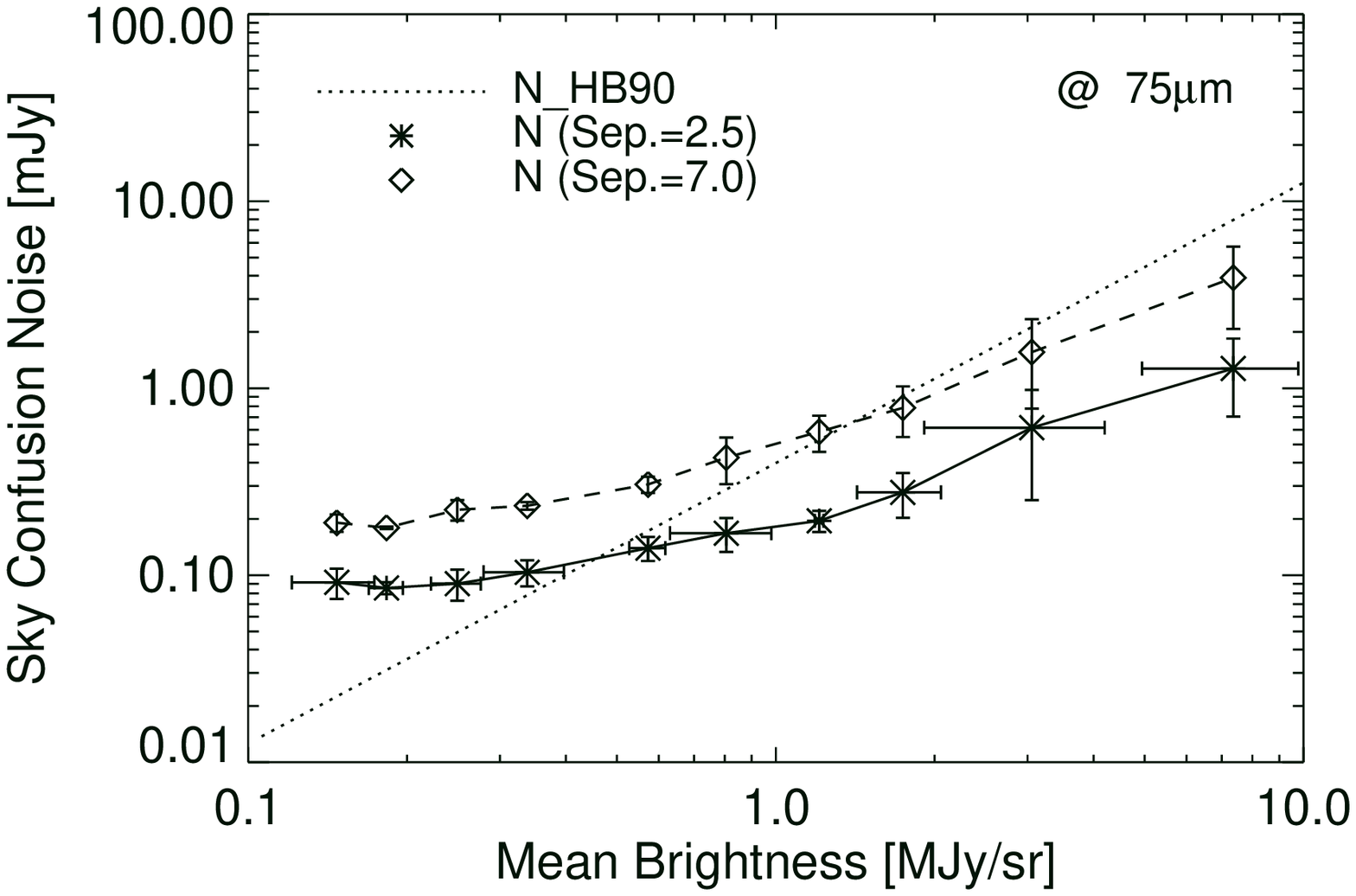, height=6.4cm}
    \psfig{figure=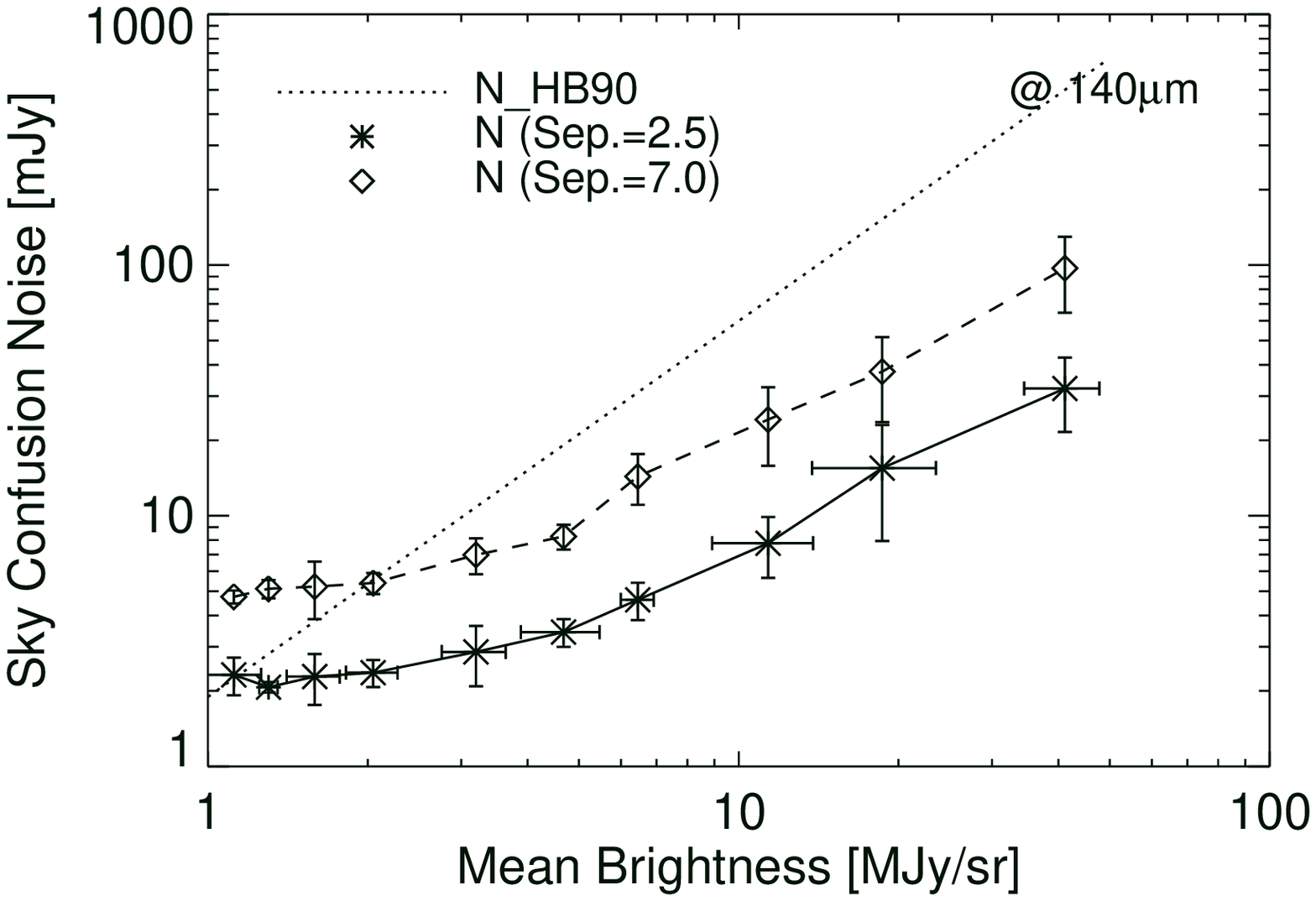, height=6.4cm}        }
   \caption{Estimated sky confusion noise for the \textit{ASTRO-F} mission. Left and right
   panels show the sky confusion noise in the WIDE-S band (75 $\mu$m) and WIDE-L band
   (140 $\mu$m), respectively. The symbols and lines are same as given in Fig.
   \ref{fig_strn_iso}.}
   \label{fig_strn_fis}
\end{figure*}

\begin{figure*}
    \centering \centerline{
    \psfig{figure=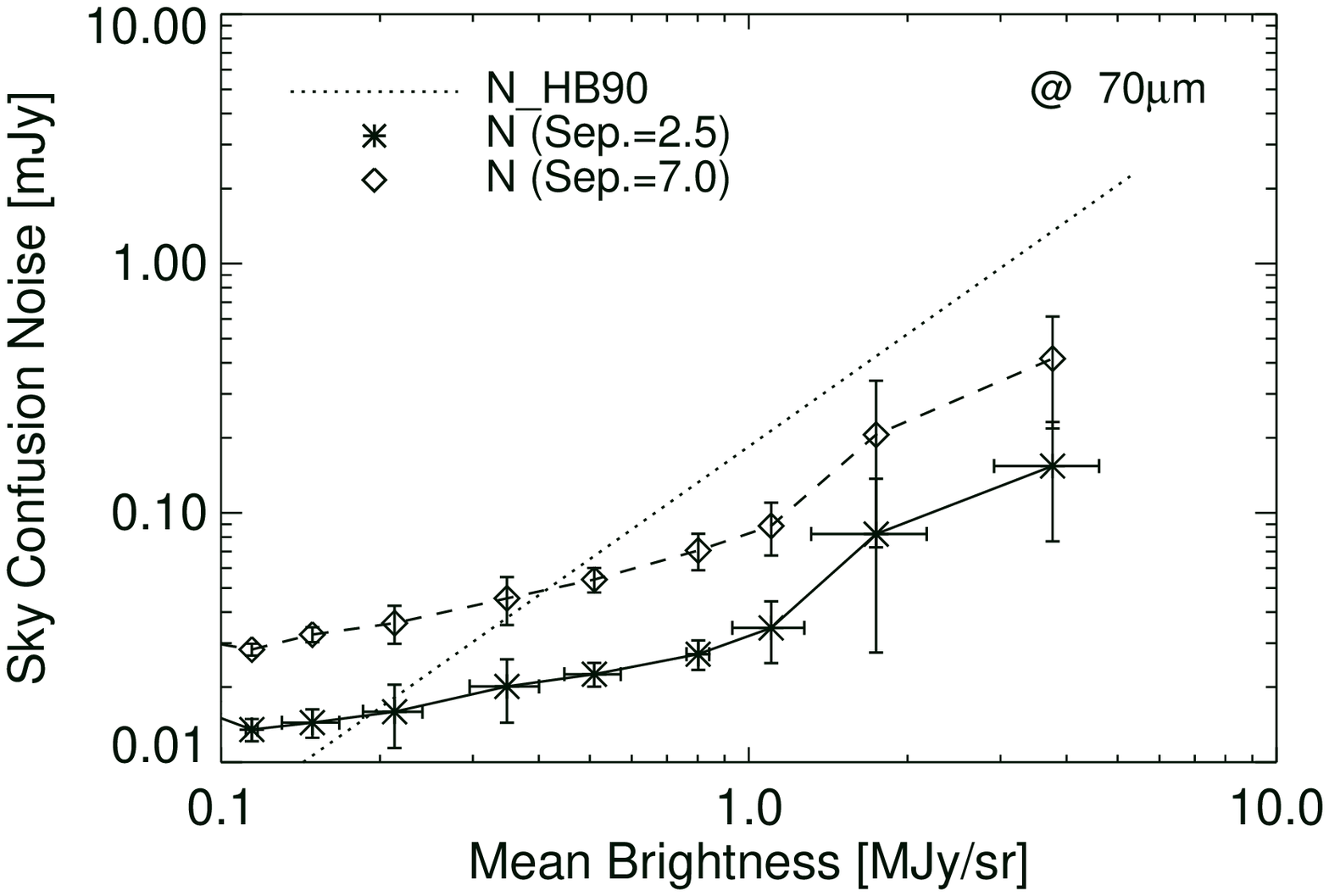, height=6.4cm}
    \psfig{figure=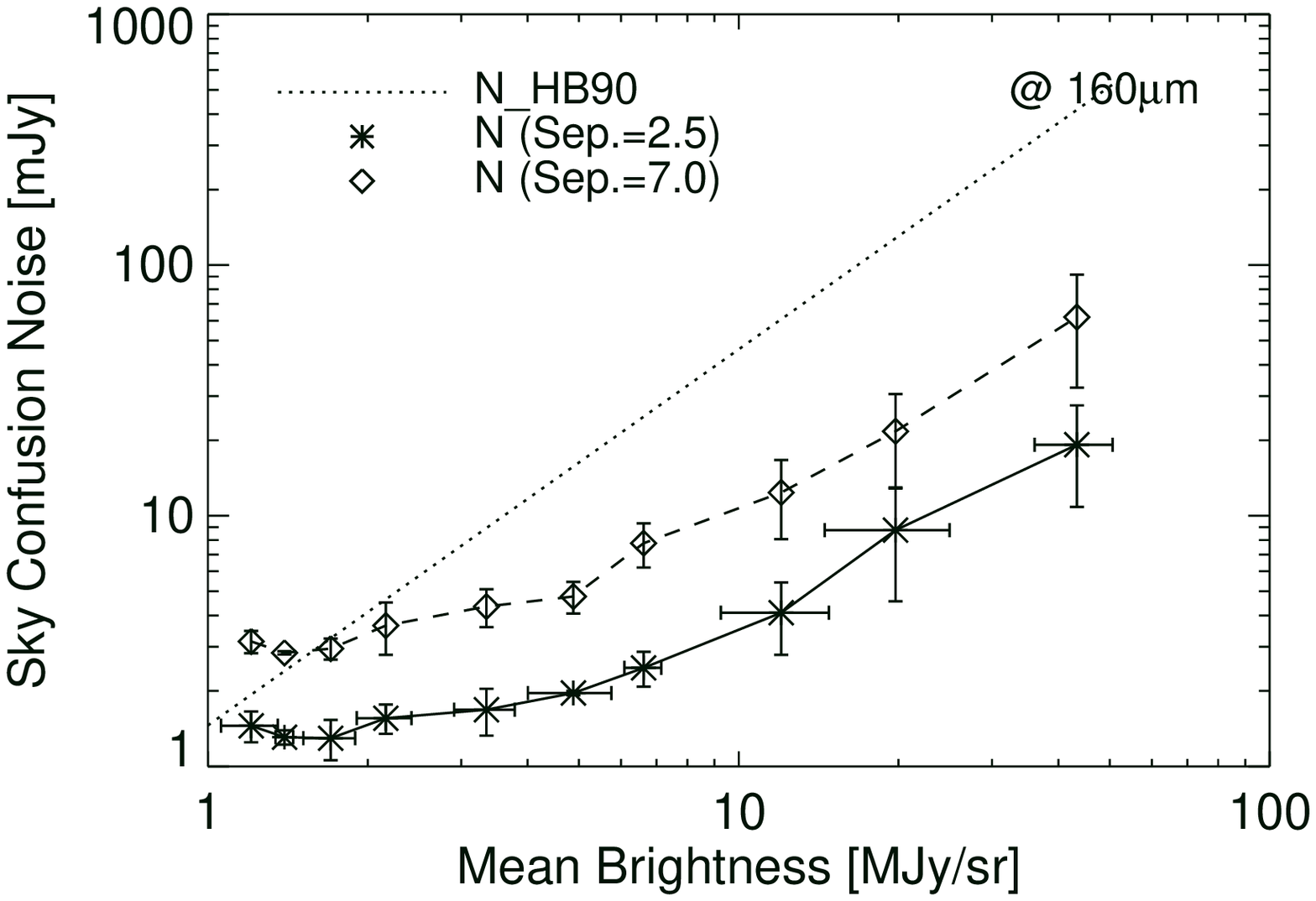, height=6.4cm}        }
   \caption{Estimated sky confusion noise for the \textit{Spitzer} mission. Left and right panels
   show the sky confusion noise for the MIPS 70 $\mu$m and 160 $\mu$m bands, respectively.
   The symbols and lines are same as in Fig. \ref{fig_strn_iso}.}
   \label{fig_strn_sirtf}
\end{figure*}

\begin{figure*}
    \centering \centerline{
    \psfig{figure=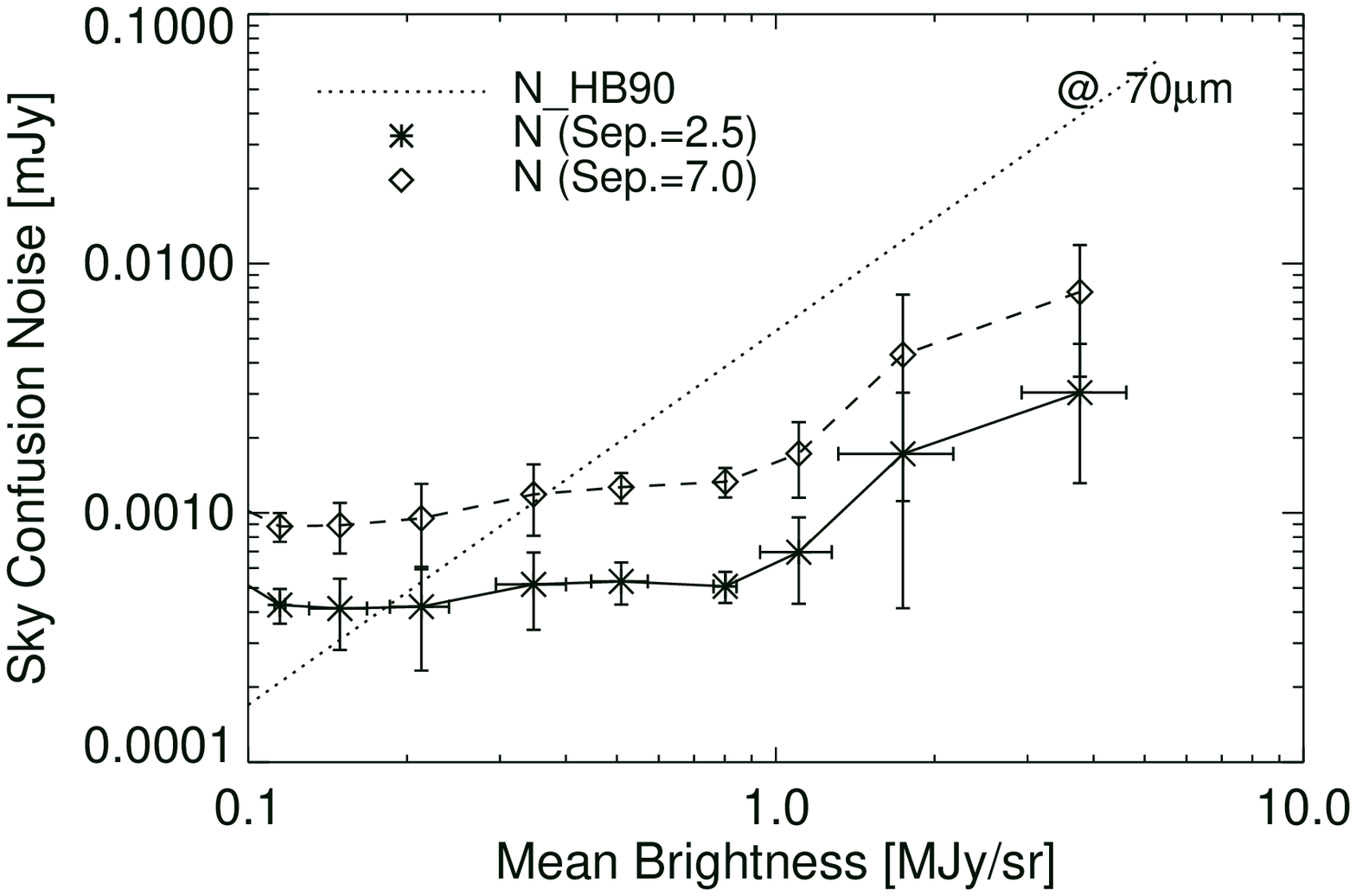, height=6.4cm}
    \psfig{figure=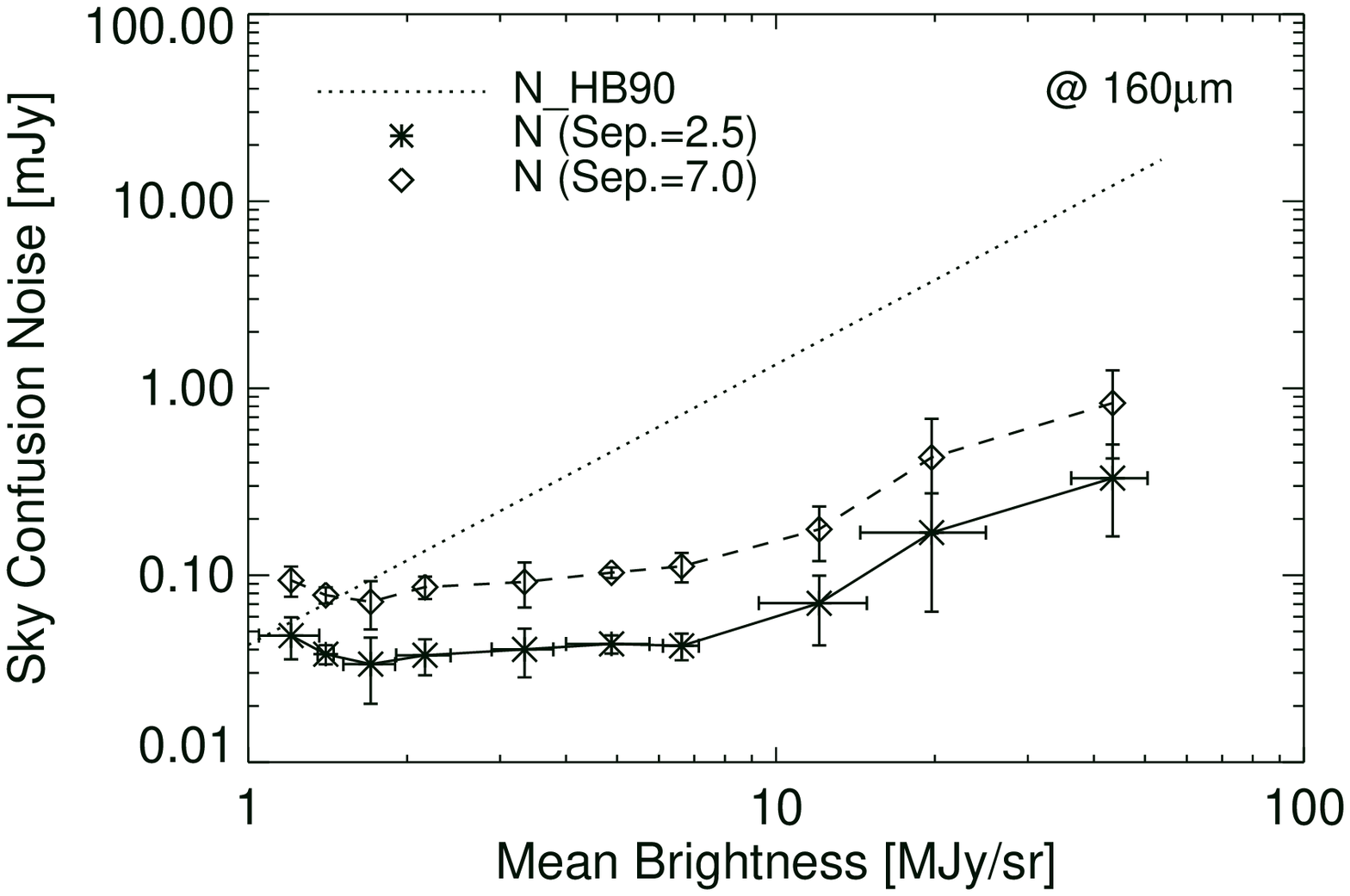, height=6.4cm}        }
   \caption{Estimated sky confusion noise for the \textit{Herschel} and \textit{SPICA} missions. Left and
   right panels show the sky confusion noise at 70 $\mu$m and 160 $\mu$m, respectively.
   The symbols and lines are same as in Fig. \ref{fig_strn_iso}.}
   \label{fig_strn_spica}
\end{figure*}

In Figs \ref{fig_strn_iso} -- \ref{fig_strn_spica}, we present our estimates of the
sky confusion noise for the \textit{ISO}, \textit{Spitzer}, \textit{ASTRO-F} and
\textit{Herschel/SPICA} space missions comparing the formula for the sky confusion
noise predicted by HB90 (hereafter HB90 formula). For \textit{ISO} results, the sky
confusion noise with $s=2.5$ is overestimated for the dark fields, but
underestimated for the bright fields (see Fig. \ref{fig_strn_iso}). With larger
separations, e.g., $s=7$, the estimated confusion noise approaches the HB90 formula
although it is still overestimated for the dark fields. We can see the same tendency
in other studies in the sky confusion noise measured from \textit{ISO} observations
\cite{herb98,kiss01}. The measured sky confusion noise for the \textit{Spitzer} and
\textit{Herschel/SPICA} missions are much lower than the predictions of HB90 except
for the dark fields (see Figs \ref{fig_strn_sirtf} and \ref{fig_strn_spica}).

\begin{figure}
    \centering \centerline{
    \psfig{figure=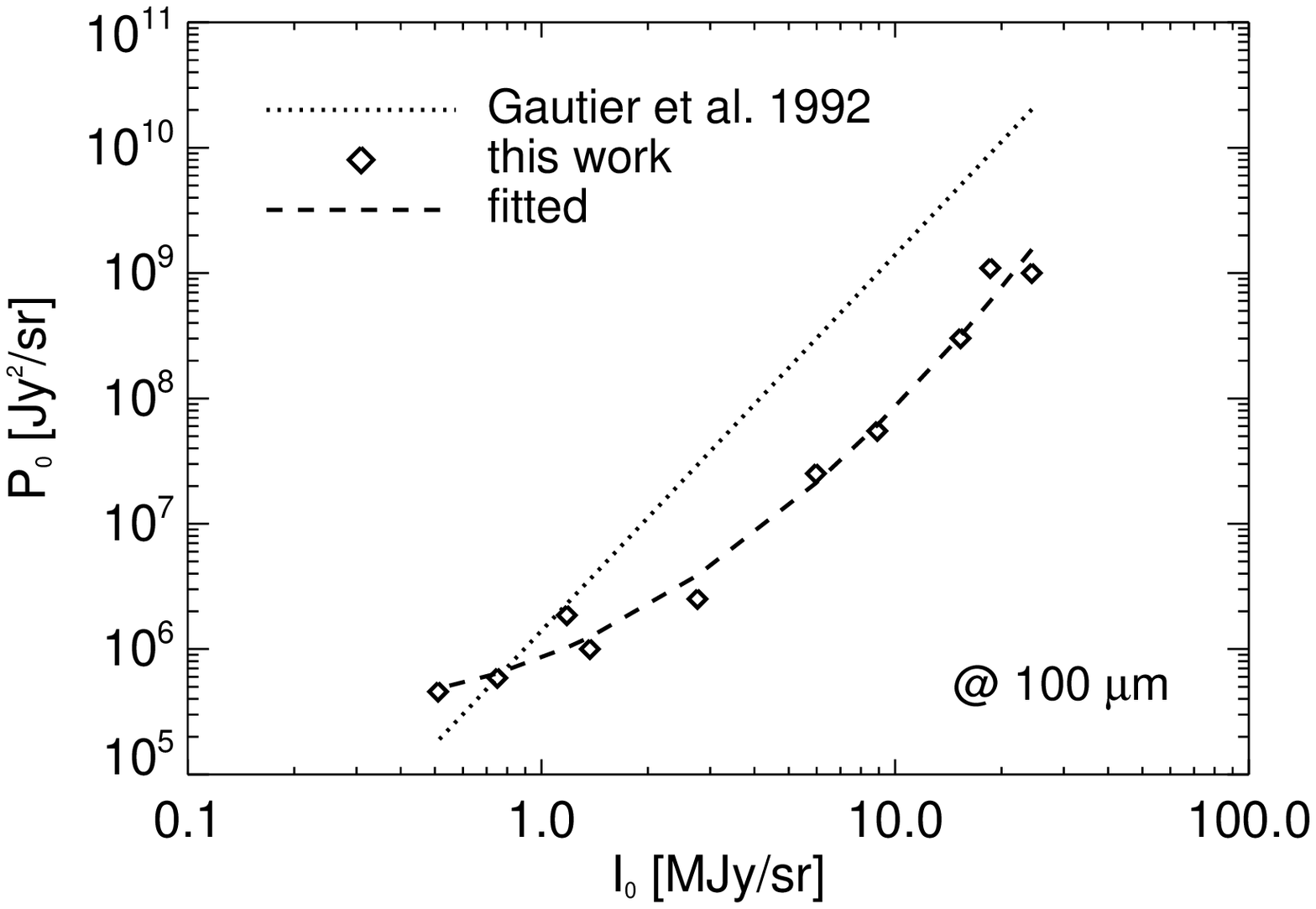, height=6.5cm}    }
   \caption{The relation between $P_0$ and $B_0^3$. The dotted line is the result
   from Gautier et al. (1992), the symbol is from our estimated $P_0$, and the
   dashed line is the fit to our result. In bright fields, values of $P_0$
   expected from Gautier et al. (1992) have higher values than those measured from our
   patches in bright fields.}
   \label{fig_rel_p0b0}
\end{figure}

Comparing the empirical relation between $P_0$ and $I_0$ by Gautier et al.
\shortcite{gautier92}, we present our estimated $P_0$ in Fig. \ref{fig_rel_p0b0}. It
shows a lower $P_0$ in bright fields and the higher $P_0$ in dark fields could cause
an underestimation in the bright fields and an overestimation in the dark fields of
the sky confusion noise. Such inconsistencies, overestimation of $P_0$ in bright
fields and underestimation of $P_0$ in dark fields, also appear in other regions of
the sky. By fitting our estimations of $P_0$, we obtained a new relation between the
$P_0$ and $I_0$. The HB90 formula assumed the wavelength dependency only through the
beam size. However, although the cirrus structure is generally preserved in other
wavelengths, the empirical relation should be scaled according to the variation of
the cirrus brightness with wavelength, i.e, cirrus spectral energy distribution.
Therefore, in order to apply our empirical formula to other wavelength bands, we
need some additional correction. For this correction, we used the ratio of the mean
brightness at the two wavelengths, e.g., $I_{160\mu\rm m}$/$I_{100\mu\rm m}$ $\sim$
2 (see Table \ref{tab_prop_patch}). For comparison with the sky confusion noise
estimated from the \textit{ISO} mission, we plot the HB90 formula to which our
empirical relation is applied (see thick dotted line in Fig. \ref{fig_strn_iso}).
Although our formula solve the discrepancies of our estimations to some extent,
there are still disagreements especially with the results for higher resolution
missions.

The HB90 formula was obtained from the analysis of the low resolution \textit{IRAS}
data at 100 $\mu$m and assumed a constant power index for the cirrus power spectrum.
In the case of the high resolution missions, since the sky confusion becomes
sensitive to the local structure rather than the large scale structure, the
calculation of  the sky confusion strongly depends upon the power spectrum estimated
for each patch and the power at the scale length corresponding to the resolution of
the detector. Therefore, we should consider carefully the combination of the
resolution and the power spectrum of the cirrus in the estimation of the sky
confusion noise. In addition, the larger discrepancy in the bright regions for the
\textit{ASTRO-F} mission compared to the prediction from \textit{ISO} observations
can be explained by an increase in the spatial resolution, although the aperture
sizes of two telescopes are similar (see the specifications of the two space
missions in Table \ref{tab_inst_para}). We conclude that the sky confusion level
predicted by the \textit{IRAS} data from which HB90 formula are derived is
significantly overestimated in the case of the higher resolution missions.

Generally the most important component superimposed on the extragalactic background
in the far-IR is the cirrus emission. However, at high spatial frequencies the
Cosmic Far-IR Background (CFIRB) fluctuations may become dominant
\cite{schl98,gui97,juvela00}. Therefore, in any estimation of the sky confusion
noise using observational data in the dark fields should consider the fluctuation
due to the CFIRB. By fitting the sky confusion noise over the mean sky brightness,
Kiss et al. \shortcite{kiss01} obtained CFIRB fluctuation of 7 $\pm$ 2 mJy at 90
$\mu$m and 15 $\pm$ 4 mJy at 170 $\mu$m. After correcting for the contribution of
the CFIRB in the estimation of the sky confusion noise, we obtained results similar
with those of Kiss et al. \shortcite{kiss01} in the dark fields (see the symbol in
circle with arrow in Fig. \ref{fig_strn_iso} at the mean brightness of $\sim$ 1.5
MJy/sr). Since the CFIRB fluctuations strongly depend upon the extragalactic source
count model, we will discuss this issue in greater detail in our forthcoming paper
[Jeong et al. 2004c \shortcite{jeong04c}, in preparation].

\subsubsection{Sky Confusion Noise for Various Separations}\label{subsec:skyconf_sep}

\begin{figure*}
    \centering \centerline{
    \psfig{figure=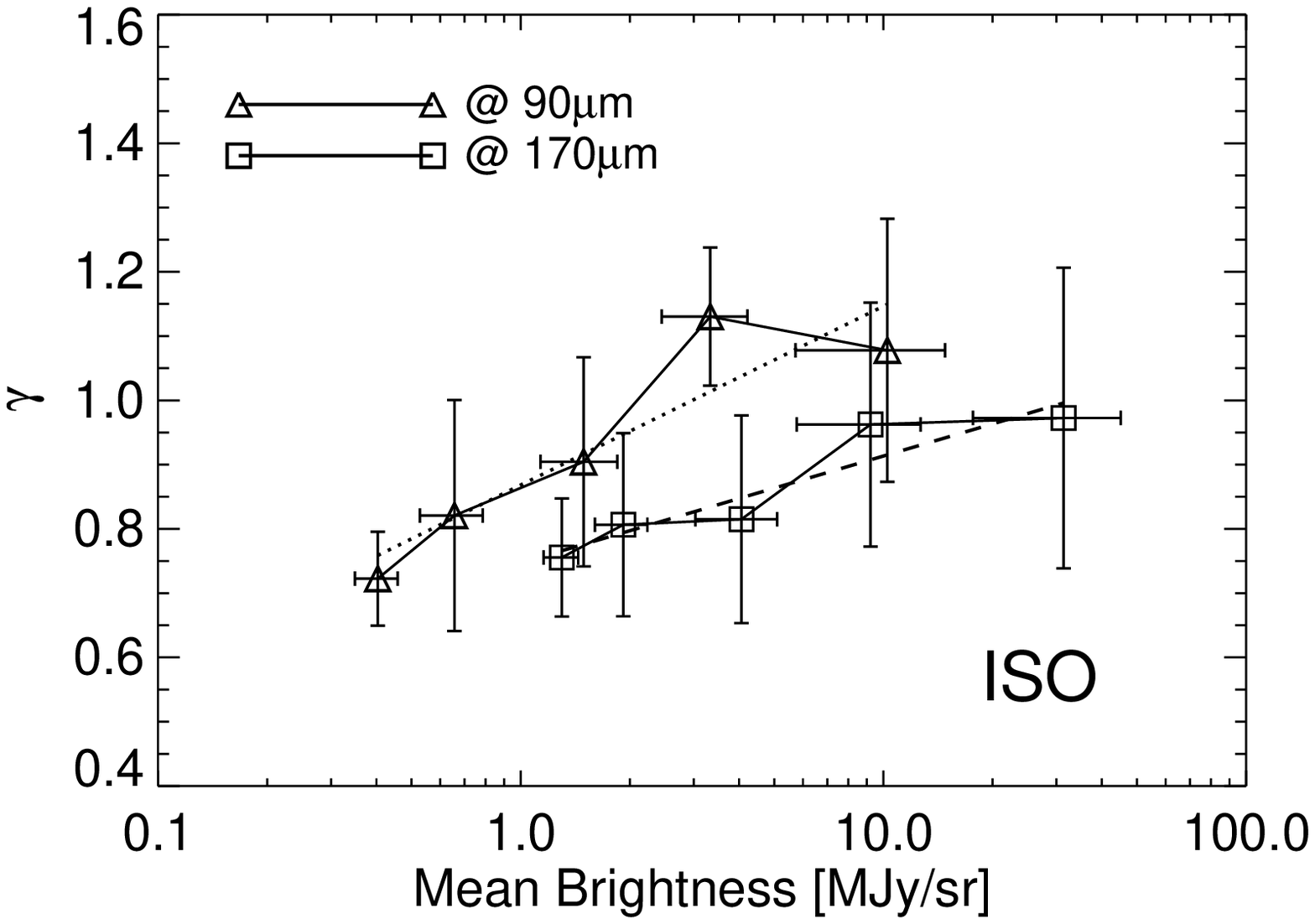, height=5.5cm}
    \psfig{figure=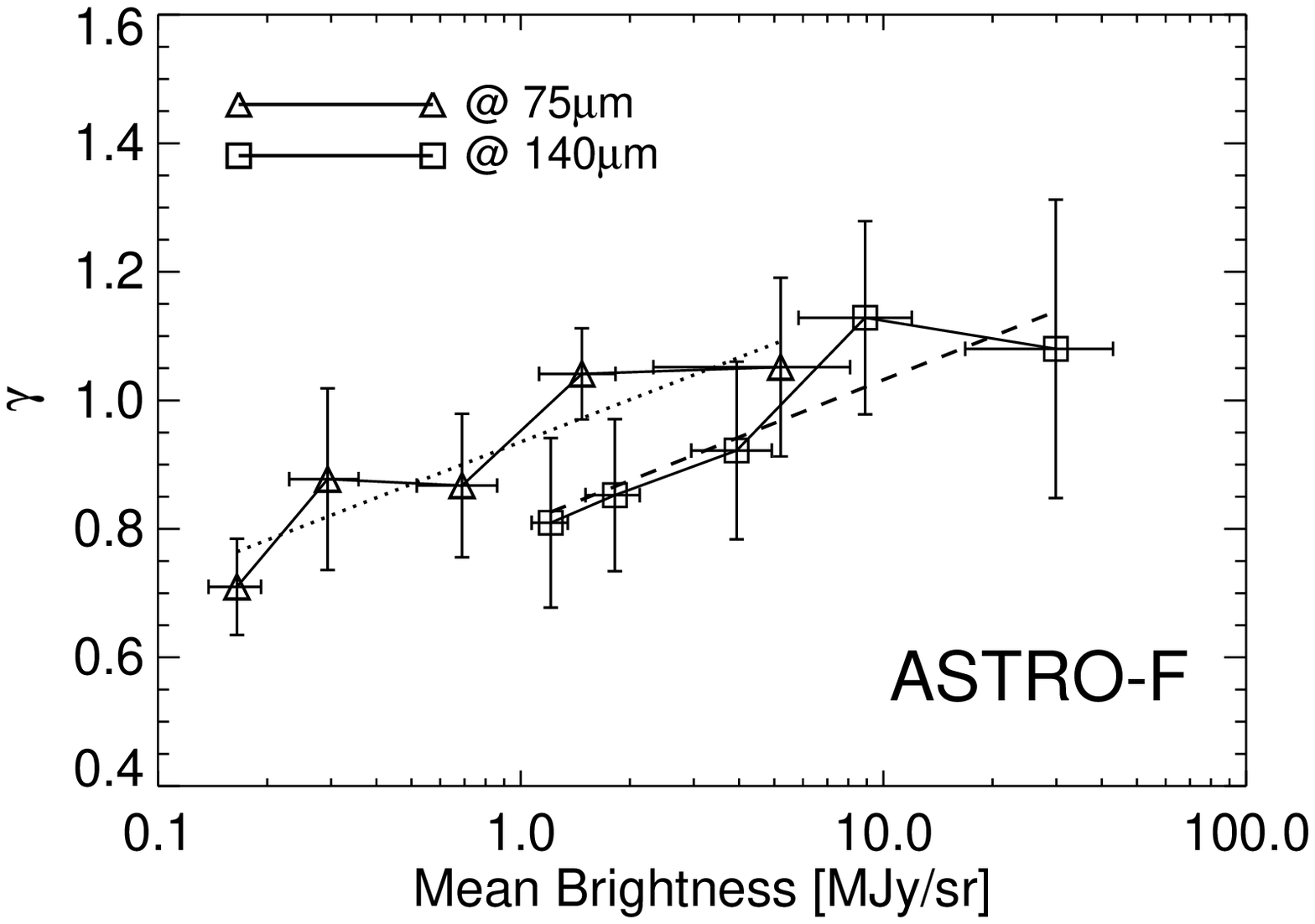, height=5.5cm} }
    \centerline{
    \psfig{figure=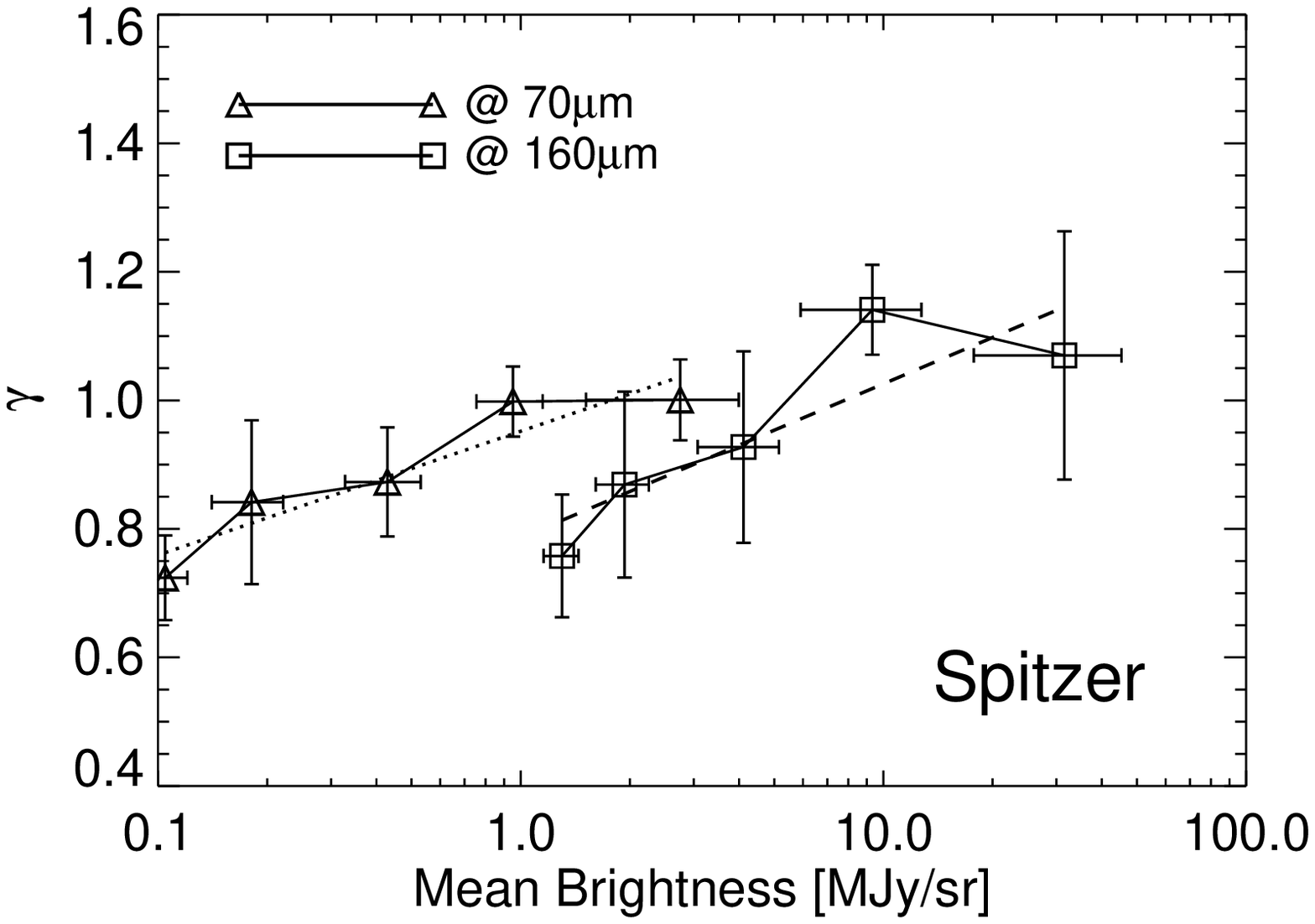, height=5.5cm}
    \psfig{figure=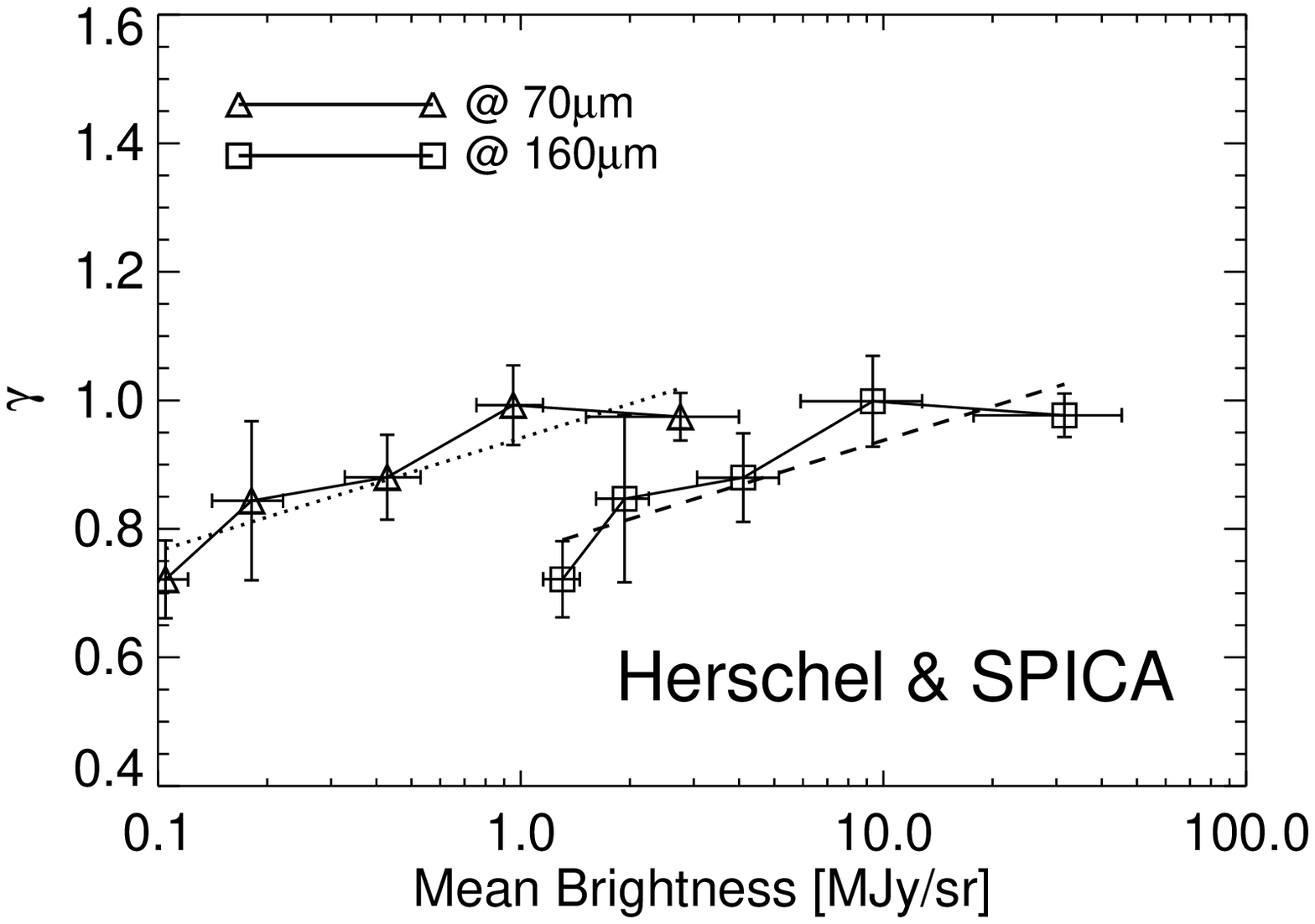, height=5.5cm}   }
   \caption{Dependency of the sky confusion noise on separation for \textit{ISO},
   \textit{ASTRO-F}, \textit{Spitzer}, \textit{Herschel} and \textit{SPICA},
   respectively. The dotted line and the dashed line is a fit to our estimation
   analysis data for SW and LW band, respectively. In the brighter regions,
   $\gamma$ has higher values than in the dark fields.}
   \label{fig_dep_sep}
\end{figure*}

Kiss et al. \shortcite{kiss01} analyzed the dependency of the sky confusion noise on
other separations by the simple power expression from \textit{ISO} observational
data:
\begin{equation}
    N(q \theta_{\rm min}) = N(\theta_{\rm min}) \times q^{\gamma}, \label{eqn_dep_sep}
\end{equation}
where $q>1$ and $\gamma$ is a constant for a specific map. We obtained $\gamma$'s
for all patches and showed $\gamma$ as a function of mean brightness for each
mission as given in Fig. \ref{fig_dep_sep}. As the sky becomes brighter, $\gamma$
becomes larger due to the prominent structure of the cirrus emission. Kiss et al.
\shortcite{kiss01} obtained a much lower $\gamma$ in dark regions, but their values
of $\gamma$ in other regions are similar to our results. This result can be
explained by two possible effects: one is that the cirrus structure observed by
\textit{ISO} is blurred by the instrumental noise in most of the dark regions and
the other is that many extragalactic point sources below the detection limit, i.e.
CFIRB fluctuations, can remove the cirrus structure. If we only consider the
component due to the cirrus in the dark fields, the values of $\gamma$ in the dark
regions by Kiss et al. \shortcite{kiss01} are similar to our results. In most of the
bright regions, the scatter of $\gamma$ shows the similar trend and this is probably
caused by the relatively large difference in the spatial structure in each region.
In the same mean brightness, $\gamma$'s in SW band are larger than those in LW band
because spatial structures should be prominent in SW band. In addition, since we use
the simulated data, changing features of $\gamma$ in two wavelength have a similar
shape. For the \textit{Herschel} and \textit{SPICA} missions, our estimations show
that $\gamma$ slowly increases and the error decreases compared with other missions,
because of the much higher resolution than the other missions considered.

\subsubsection{Effect of Power Index $\alpha$}

In this study, we assume that the structure of cirrus is independent of wavelength.
However, recent papers reported on enhanced dust emissivity at some medium-to-high
density clouds in LW band of Far-IR due to the presence of a cold dust component (T
$\leq$ 15K) \cite{cam01,delbur03,step03}. This result imply that the cirrus
structure can be changed in LW band. Kiss et al. \shortcite{kiss03} suggested that
the power index of the power spectrum also depends upon both wavelength and surface
brightness due to the coexistence of dust components with various temperatures
within the same field and cold extended emission features (usually, $-2.0 < \alpha <
-4.0$). Using the assumption that the sky confusion noise is proportional to the
scale length (see equation \ref{eqn_rel_strps}), we can estimate the sky confusion
for different power indices. The ratio $\psi$ of the sky confusion noise with the
power index of $\alpha + \epsilon$ to that with the power index of $\alpha$ can be
defined as
\begin{equation}
    \psi = \frac{N(\alpha + \epsilon)}{N(\alpha)}, \label{eqn_scn_index}
\end{equation}
where $\epsilon$ is the contribution to the power index from any other structure in
the power spectrum. In this calculation, we fix the power at the scale length of the
resolution limit of the map ($\sim$ 6.1 arcmin) and wavelength at 100 $\mu$m from
the assumption that the power over this scale is not affected by the extra
components proposed by Kiss et al. \shortcite{kiss03}. Table \ref{tab_scaled_strn}
lists the ratio of the sky confusion noise for the different power indices for each
space mission covering power indices of the power spectrum on the cirrus emission.
Since the fluctuation at smaller scales is more sensitive to the power index, the
sky confusion noise is much more dependent upon the power index for the space
missions with higher resolutions. As seen in Table \ref{tab_prop_patch}, our
estimated power indices in the bright regions ($\alpha$ $>$ 3.3) are somewhat higher
than those in low density regions ($\alpha$ $<$ 2.8). From the recent
\textit{Spitzer} observation, Ingalls et al. \shortcite{ingalls04} obtained the
power index of -3.5 at 70 $\mu$m in the Gum Nebula. Therefore, if this varying power
index is not so large, it does not affect severely the final sensitivity values.

\begin {table}
\centering \caption {Ratio $\psi$ of the sky confusion noise for the different power
indices.} \label{tab_scaled_strn} \vspace{5pt}
\begin{tabular}{@{}ccccc}
\hline\vspace{-5pt} \\
& $\epsilon$~$^a$ = -1.0 \span\omit & $\epsilon$ = 1.0 \span\omit \vspace{5pt} \\
Space Mission & SW & LW & SW & LW \vspace{5pt}
\\\hline \vspace{-10pt}
\\ \textit{ISO} & 0.13 & 0.19 & 1.7 & 1.2 \vspace{5pt}
\\ \textit{Spitzer} & 0.083 & 0.12 & 2.8 & 1.9 \vspace{5pt}
\\ \textit{ASTRO-F} & 0.10 & 0.13 & 2.2 & 1.8 \vspace{5pt}
\\ \textit{Herschel}  & 0.041 & 0.061 & 5.6 & 3.8 \vspace{5pt}
\\  \textit{SPICA} & 0.041 & 0.061 & 5.6 & 3.8 \vspace{5pt}
\\ \hline
\end{tabular}
\medskip
\begin{flushleft}
{\em $^a$} contribution index in the power spectrum.
\end{flushleft}
\end{table}

\section{PHOTOMETRIC MEASUREMENTS OF SKY CONFUSION NOISE}\label{sec:scn_phot}

In Section \ref{sec:stat_analy_scn}, we estimated the sky confusion noise by the
fluctuation analysis. The sky confusion noise should affect the source detection
efficiency, causing a deterioration in the detection limit. In this section, we
obtain the measured sky confusion noise by carrying out photometry on realistically
simulated data.

\subsection{Source Distribution}\label{sec:source_dist}

The distribution of sources per unit area on the sky can be described as a function
of the flux density and depends upon both the spatial distribution of the sources
and their luminosity function. For simplicity, we assume the number of sources whose
flux is greater than flux $F$, $n(>F)$, is a power-law function of $F$,
\begin{equation}
n(>F) = n_0 (> F_0) \left({F\over F_0}\right)^{-\omega}, \label{eqn_sdist}
\end{equation}
for $F_{\rm min} < F < F_{\rm max}$, where $n_0$ and $F_0$ are normalization
constants for number of sources and for flux, respectively, $F_{\rm min}$ is the
minimum flux, $F_{\rm max}$ is the maximum flux in the source distribution.

\begin{figure}
    \centering \centerline{
    \psfig{figure=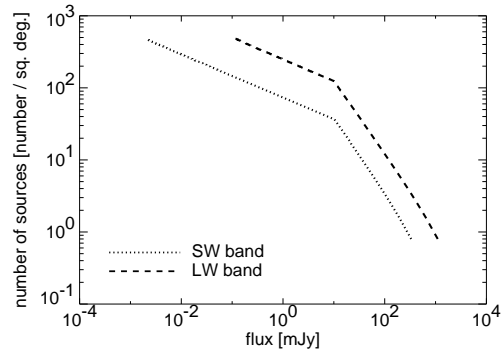, height=5cm} }
   \caption{Source distribution in the SW band and LW band. We use different slopes
   ($\omega$ = 1.0 and $\omega$ = 0.3) for the power law source distribution at the
   boundary flux of 10 mJy in order to reduce the effect of the source confusion.}
   \label{fig_source_dist}
\end{figure}

The source confusion caused by the overlapping of adjacent sources mainly depends
upon the source distribution and the beam profile \cite{cond74,fran89}. Source
confusion becomes important as the observation sensitivity increases since there are
usually more faint sources than brighter ones. Currently favorable source count
models require strong evolution in order to fit the \textit{ISO} data from mid- to
far-IR, the SCUBA data at sub-mm wavelengths, and the Cosmic Infrared Background
(CIRB) at 170 $\mu$m
\cite{oliver97,smail97,kawara98,hugh98,aussel99,puget99,esf00,serjeant00,lagache00,mat00,scott02}.
In our study, we use a simple source distribution for the purpose of investigating
only the effect of the sky confusion. We will discuss the source confusion with more
realistic source count models in the forthcoming paper. In order to avoid the
contributions from any source confusion itself, we assume rather sparse distribution
of sources. However, the estimate of detection limit becomes rather uncertain, if
there are too few sources. Therefore, we have employed a model for the $n(F)$
utilizing a distribution with two slopes, $\omega$ = 1.0 for bright flux region and
$\omega$ = 0.3 for faint flux region (see Fig. \ref{fig_source_dist}), in order to
derive an accurate value for the sky confusion limits without source confusion
effect. Since the sky confusion noises in the SW bands are much lower than those in
the LW bands, we set different normalization constants and minimum flux values
$F_{\rm min}$, i.e., $F_{\rm min}$ = 0.001 mJy and $n_0 (> F_0)$ = 3 in the SW band,
$S_{\rm min}$ = 0.1 mJy and $n_0 (> F_0)$ = 10 in the LW band, where $F_0$ is set to
be 100 mJy (see Fig. \ref{fig_source_dist}).

\subsection{Source Detection}\label{sec:sub_source_det}

\begin{figure*}
    \centering \centerline{
    \psfig{figure=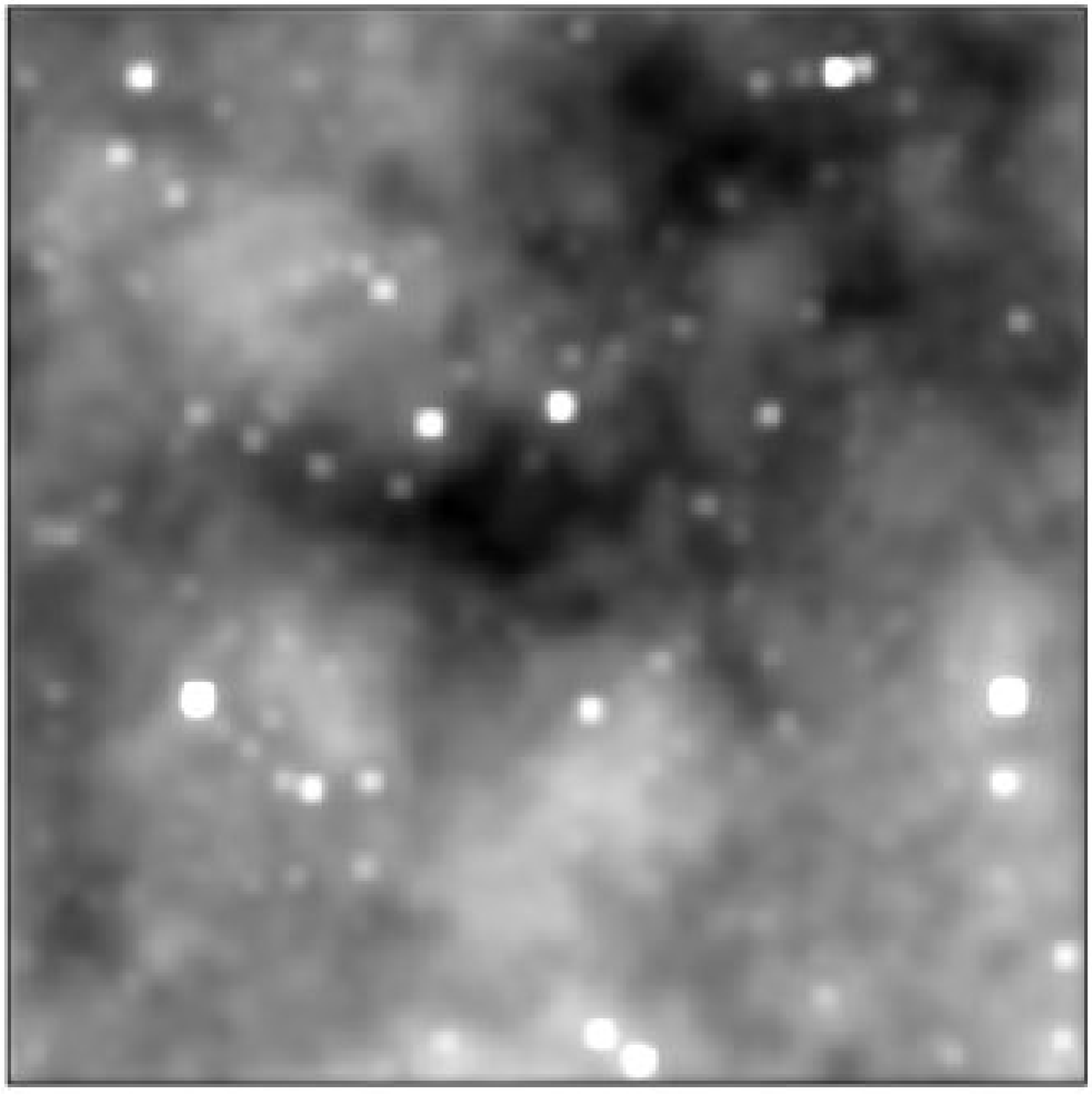, height=6cm}
    \psfig{figure=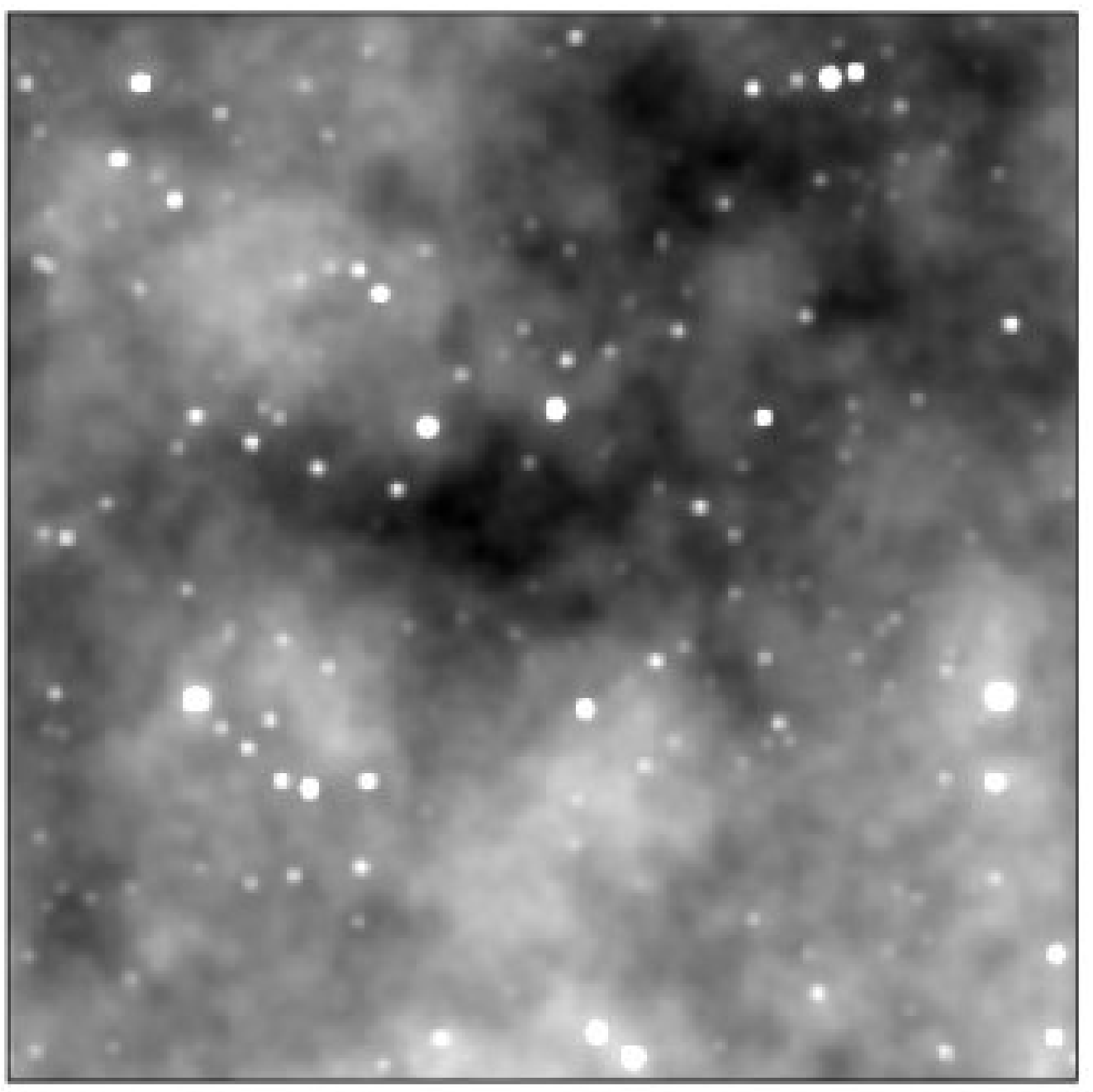, height=6cm}   }
    \centerline{
    \psfig{figure=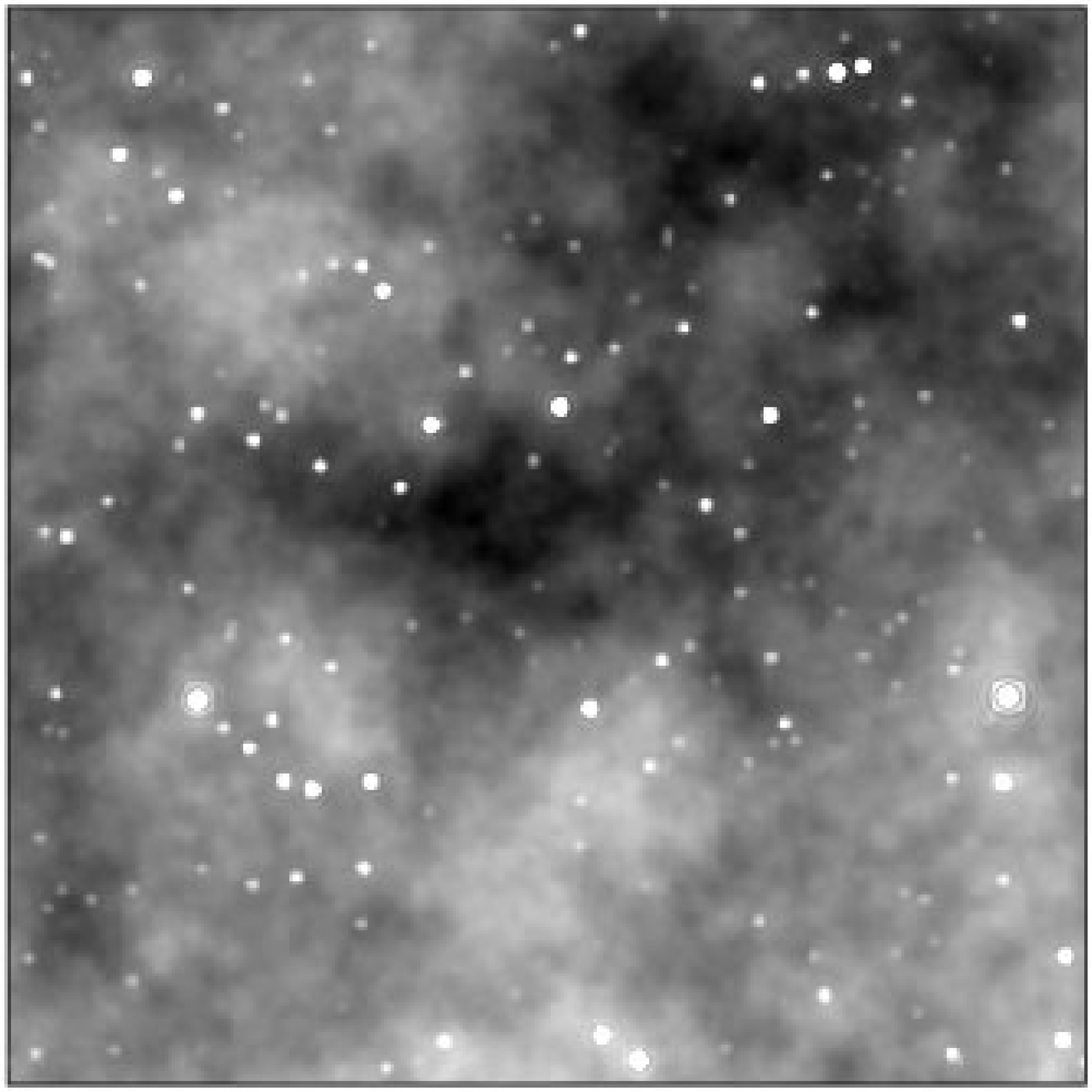, height=6cm}
    \psfig{figure=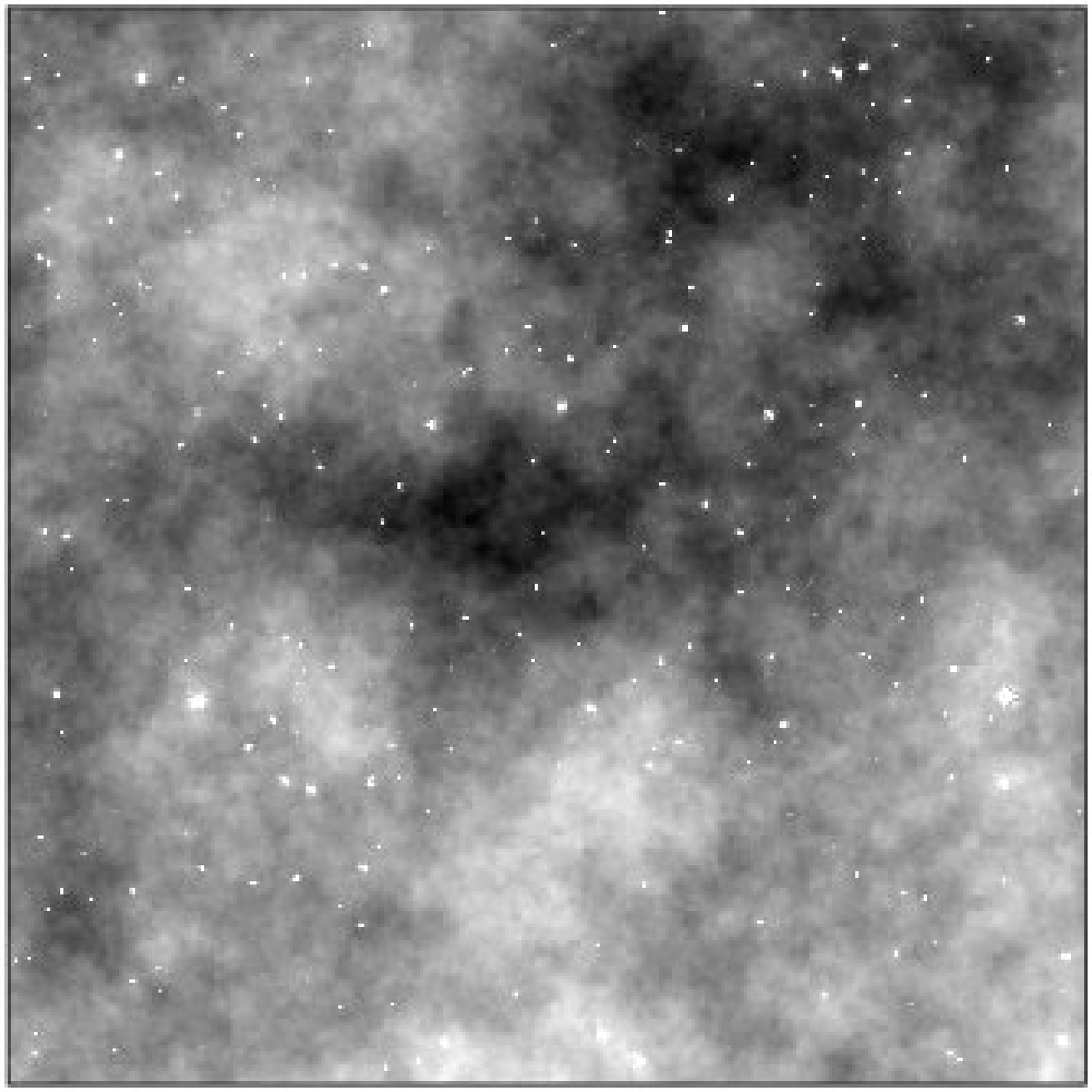, height=6cm}   }
   \caption{Simulated images including point sources in the LW band for \textit{ISO}
   (upper-left), \textit{ASTRO-F} (upper-right), \textit{Spitzer} (lower-left), \textit{Herschel}
   and \textit{SPICA} (lower-right) missions. The mean brightness of the cirrus background is
   2 MJy~sr$^{-1}$ at 160 $\mu$m.}
   \label{fig_sim_img}
\end{figure*}

We generate images including point sources convolved with the beam profile of each
mission using the source distribution described in Section \ref{sec:source_dist}.
Fig. \ref{fig_sim_img} shows the simulated images for the various missions
considered. As the detector pixel and the beam profile become smaller, more sources
and smaller structure in the cirrus emission appear.

\begin{figure*}
    \centering \centerline{
    \psfig{figure=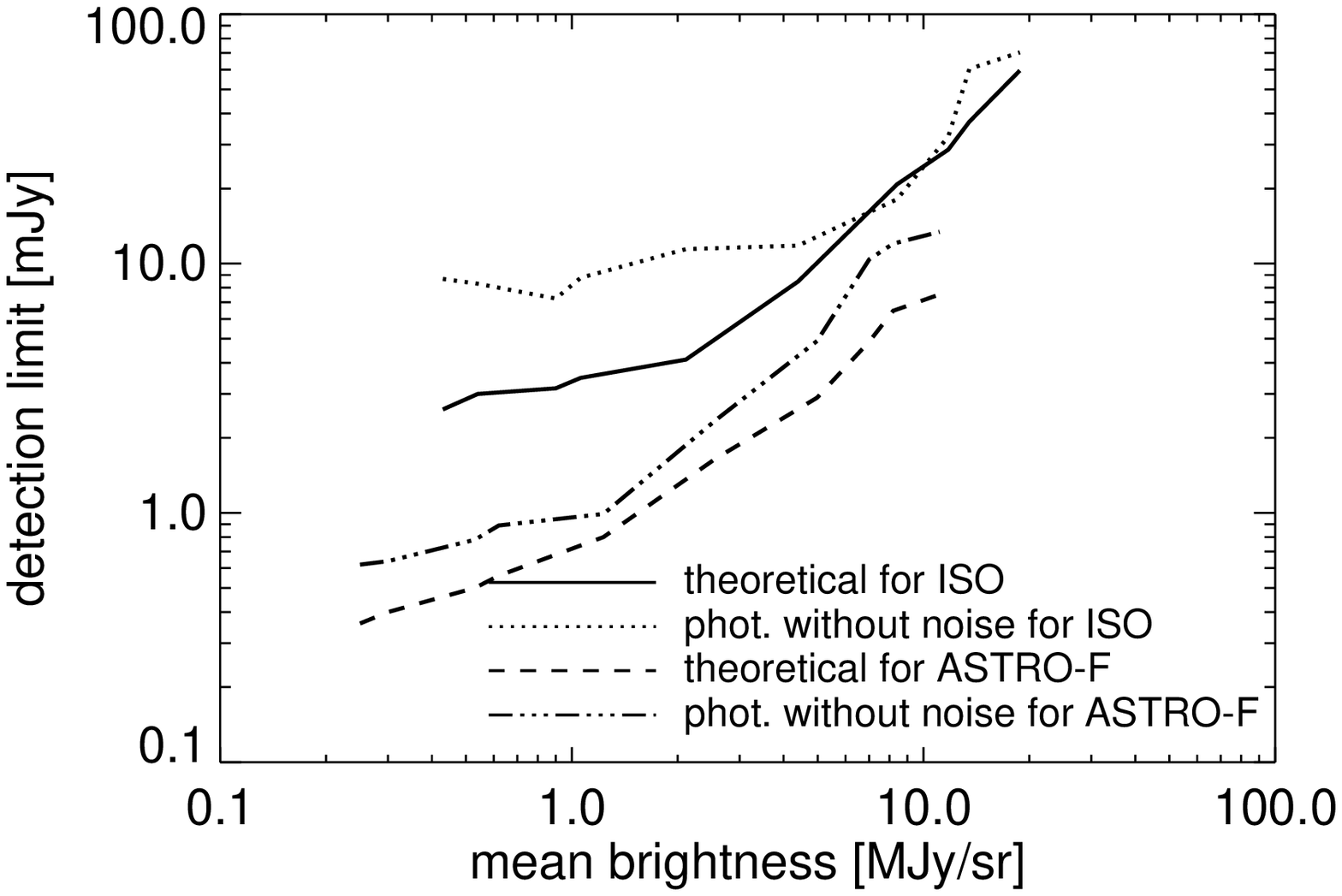, height=5.5cm}
    \psfig{figure=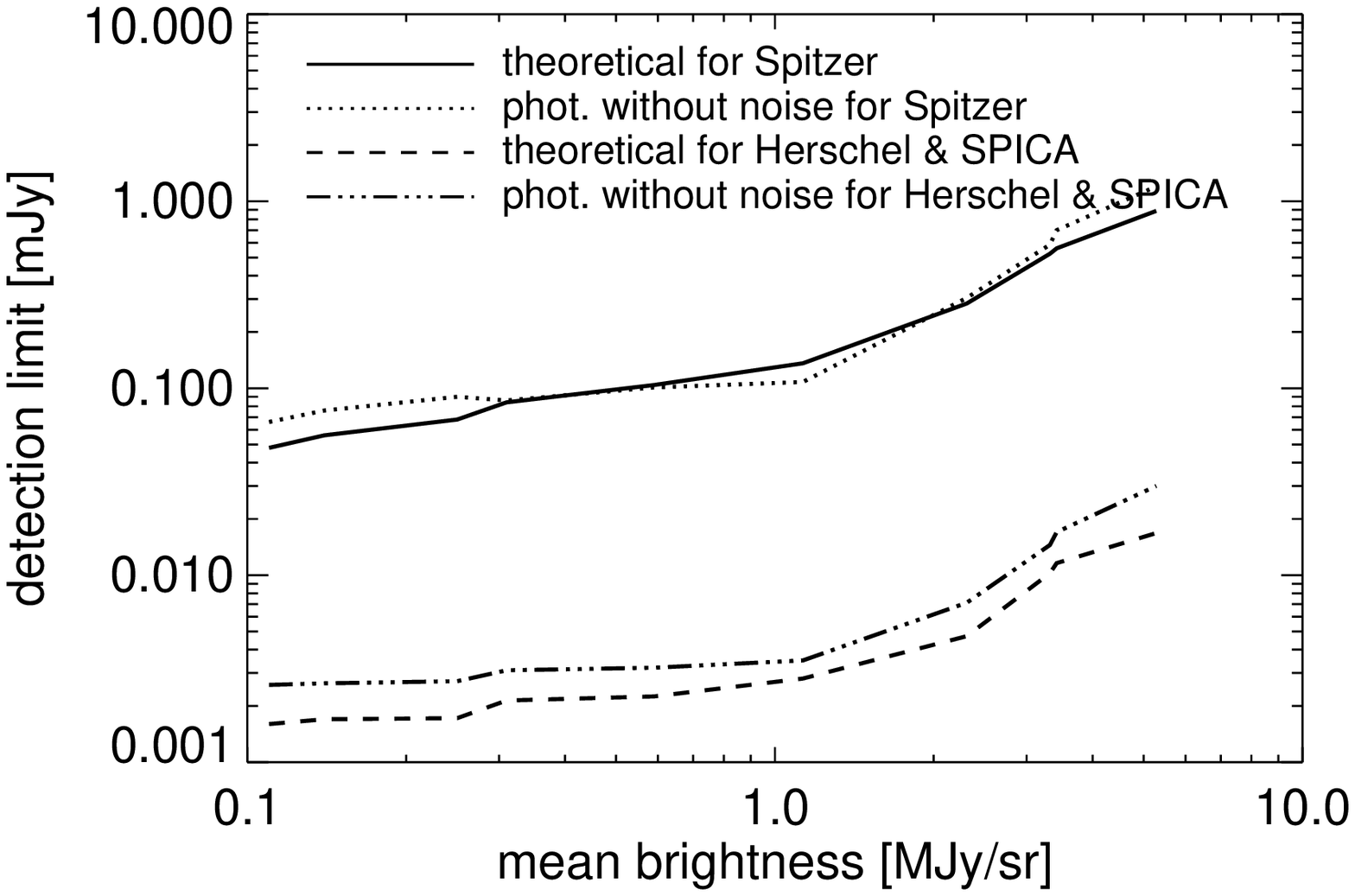, height=5.5cm}   }
    \centerline{
    \psfig{figure=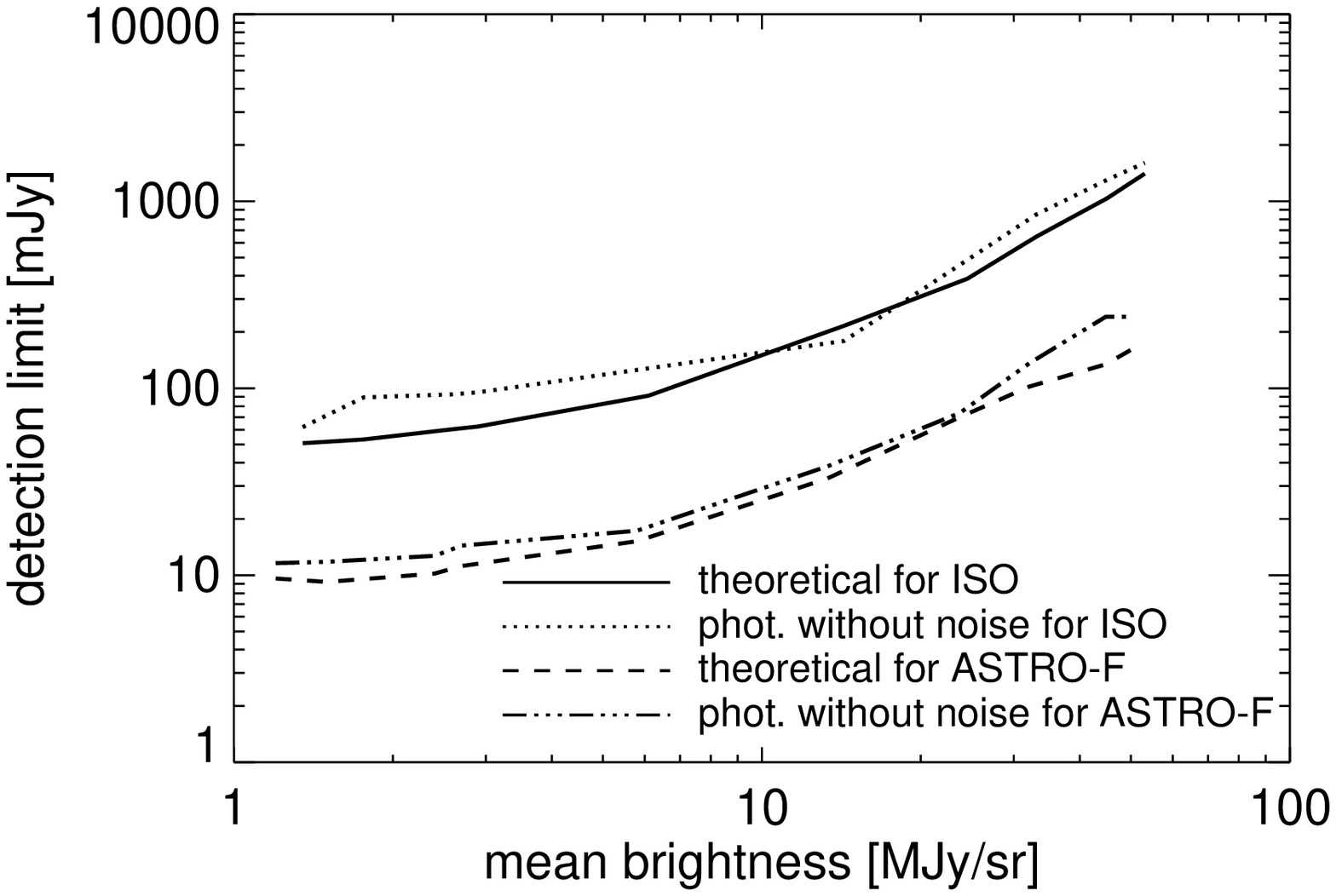, height=5.5cm}
    \psfig{figure=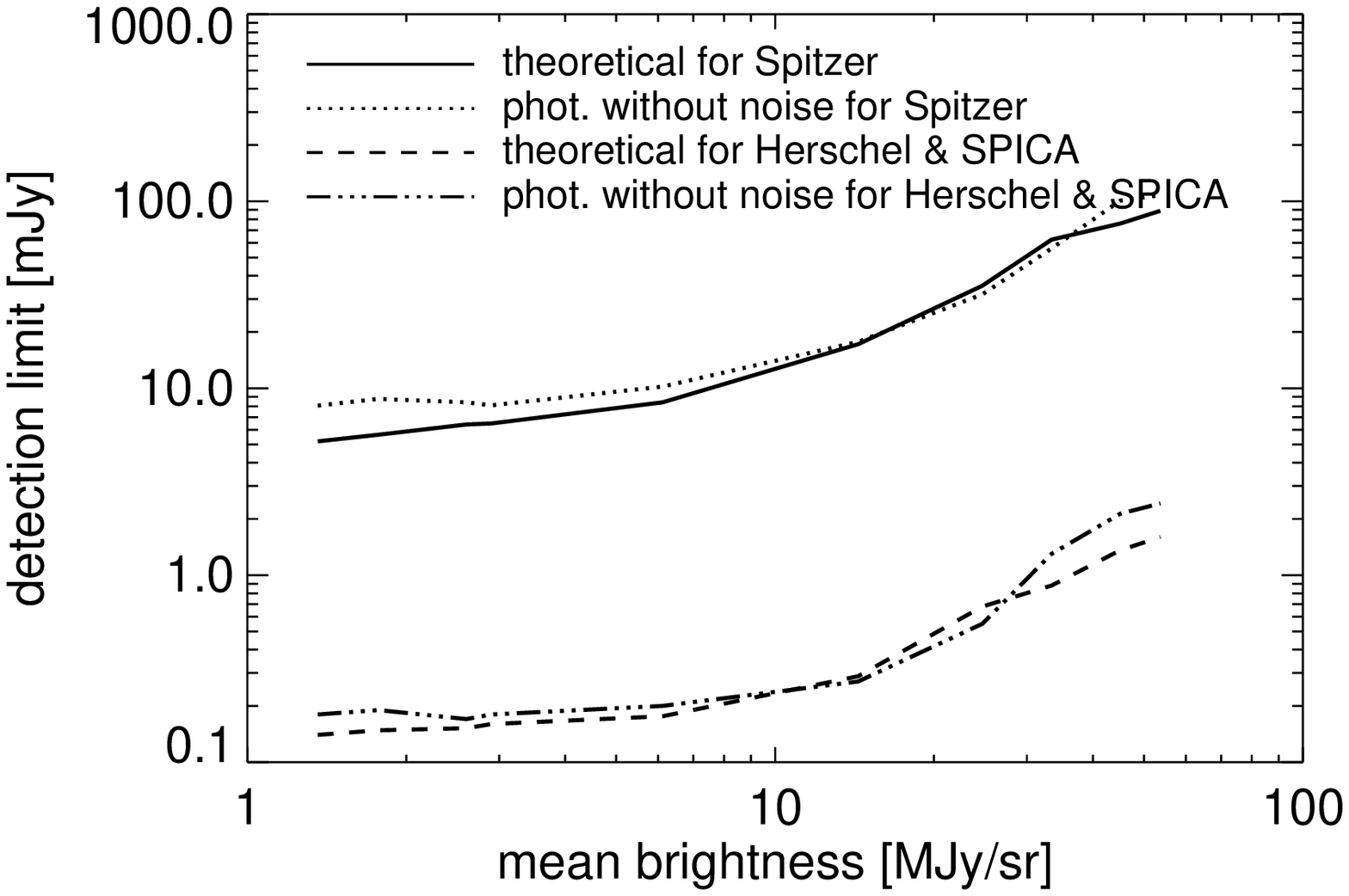, height=5.5cm}   }
   \caption{Estimated detection limit by photometry.
   Figures show the detection limit and 4 times sky confusion noise estimated
   from the fluctuation analysis for the \textit{ISO} and \textit{ASTRO-F}
   missions (left) and \textit{Spitzer}, \textit{Herschel} and \textit{SPICA}
   missions (right). Upper and lower panels show the results for the SW band and LW band, respectively.}
   \label{fig_strn_phot}
\end{figure*}

We carried out aperture photometry on the simulated images using the SExtractor
software \textit{v}2.2.2 \cite{bert96}. There are several parameters to be fixed to
perform the photometry, but the most influential parameters are the size of a
background mesh for estimating background level and the threshold for the source
detection in this aperture photometry. In order to optimise for better reliability
of the detected sources and reducing the rate of false detection, we make trials by
changing two parameters. Finally, we set the size of the background mesh to be 2.5
times of the measuring aperture, and the detection threshold as 4$\sigma$. The final
detection limit is determined by the minimum flux of detected point sources. We
found that the detection limits determined from 4$\sigma$ criteria are consistent
with the 4 times of sky confusion noise measured from the fluctuation analysis. Note
that our sky confusion noise estimated from the fluctuation analysis is a 1$\sigma$
fluctuation. In Fig. \ref{fig_strn_phot}, we compare the detection limit by
photometry with the sky confusion noise for each mission. For the \textit{ISO} and
\textit{ASTRO-F} missions, the results from photometry give relatively higher
detection limits than the theoretical estimations via fluctuation analysis. This
trend results from the larger detector pixel size compared to the FWHM of the beam
profile. The large detector pixel size of the \textit{ISO} mission significantly
degraded the performance of the detection of the point sources (e.g., the left
panels in Fig. \ref{fig_strn_phot}).

\section{SUMMARY AND DISCUSSION}\label{sec:summary}

Based on the observed 100 $\mu$m dust map and the models of a dust spectrum, we
generated high resolution background maps at wavelengths ranging from 50 to 200
$\mu$m. Using these simulated cirrus maps, we estimated the sky confusion noise for
various IR space missions such as \textit{ISO}, \textit{Spitzer}, \textit{ASTRO-F},
\textit{Herschel} and \textit{SPICA}. Since we have the observational results only
from \textit{ISO}, we compared the results of our simulation with the \textit{ISO}
data. We found that the sky confusion noise estimated with our simulated maps are
consistent with the \textit{ISO} results. However, in the dark fields the sky
confusion noise is more weakly dependent upon the beam separation parameter than in
the bright fields in the case of the \textit{ISO} observation. We conclude that this
is due to the fact that the instrumental noise dominates in the dark regions or
alternatively, the CFIRB fluctuation is more important. We also found that the sky
confusion predicted from the \textit{IRAS} data is significantly overestimated in
the case of the large aperture telescopes, except for the dark fields.

We have confirmed our results through a realistic simulation. We performed
photometry on simulated images including point sources with a sparse source
distribution in order to avoid the effects of confusion due to crowded point
sources. The detection limits obtained from the photometric analysis agree with the
sky confusion noise estimated using fluctuation analysis except for \textit{ISO} and
\textit{ASTRO-F}. The discrepancies for these missions are due to the large detector
pixel size compared to the FWHM of the beam size.

\begin{figure}
    \centering \centerline{
    \psfig{figure=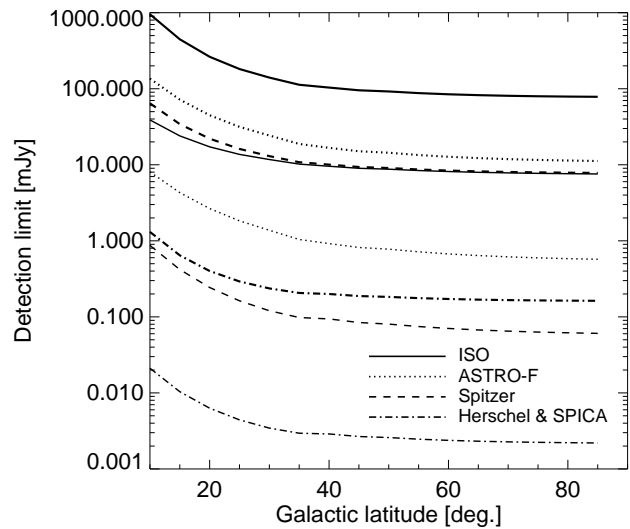, width=8.5cm}    }
    \vspace{25pt}
   \caption{Detection limits due to the Galactic cirrus as a function of Galactic latitude.
   The two line plotted for each mission are for the SW band (lower line) and the  LW
   band (upper line).}
   \label{fig_detlim_gb}
\end{figure}

The mean brightness of the cirrus emission usually decreases with increasing
Galactic latitude \cite{boul88}. In order to estimate the detection limits as a
function of Galactic latitude, we derived a simple formula for each wavelength band.
Because the cirrus emission is extremely strong near the Galactic centre, we
excluded the Galactic latitudes $|b|<10^\circ$. Fig. \ref{fig_detlim_gb} shows the
detection limits as a function of Galactic latitude. The detection limits for all
missions appear to saturate beyond $b \sim 30^\circ$.

\begin{figure}
    \centering \centerline{
    \psfig{figure=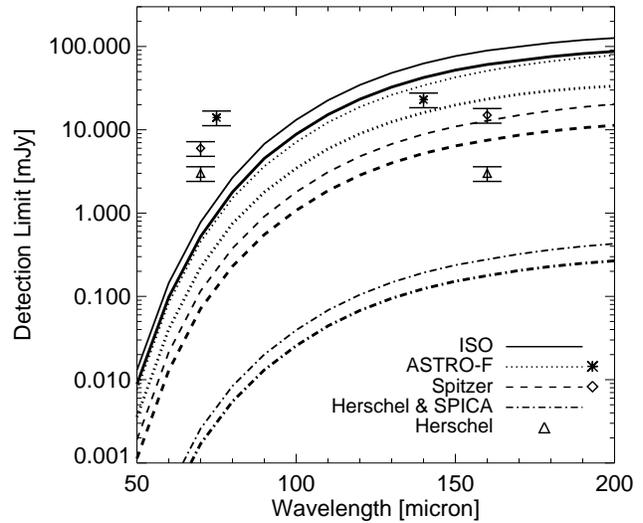, width=8.5cm}    }
    \vspace{25pt}
   \caption{Detection limits due to Galactic cirrus at mean and low sky brightness
   in each band. The mean sky brightness in the SW and LW bands is set to
   1 MJy~sr$^{-1}$ and 15 MJy~sr$^{-1}$, respectively. The lower value for each
   detection limit corresponds to the detection limit at low sky brightness usually at
   high Galactic latitudes. The symbol shows the 5$\sigma$ sensitivity for the
   \textit{ASTRO-F}, \textit{Spitzer}, \textit{Herschel} missions without confusion
   and the error bar corresponds to 1$\sigma$ sensitivity.}
   \label{fig_detlim_scn}
\end{figure}

Fig. \ref{fig_detlim_scn} summarises the final detection limits for point sources at
mean and low sky brightness regions due to the Galactic cirrus. In addition, we also
plot the currently estimated 5$\sigma$ detection limits for sources of each mission.
The detection limits only take into account the instrumental noise. The instrumental
noise for \textit{ASTRO-F} mission is explained in detail in Jeong et al. (2003;
2004a; 2004b). The integration time is 500 sec for the \textit{Spitzer} mission
(\textit{Spitzer} Observer's Manual\footnote{Further information can be found at the
following url: \it{http://ssc.spitzer.caltech.edu/mips/sens.html}}) and 1 hour for
the \textit{Herschel} mission \cite{pilb03}. As shown in Fig. \ref{fig_detlim_scn},
sky confusion almost approaches the detection limit in the LW band of the
\textit{ASTRO-F} and \textit{Spitzer} missions. Although the sky confusion does not
severely affect the detection limits of \textit{Herschel} mission, it can affect the
detection limit of the \textit{SPICA} because it will have a large aperture
telescope cooled to very low temperatures in order to achieve exceptional
sensitivity in the far-IR (see Nakagawa 2004 for the detailed information of the
\textit{SPICA} mission).

\section*{Acknowledgment}
This work was financially supported in part by the KOSEF Grant R14-2002-058-01000-0.
Chris Pearson acknowledges a European Union Fellowship to Japan. We thank Kyung Sook
Jeong for careful reading of the manuscript and fruitful suggestions.

\end{document}